# Predictive Analytics for Collaborators' Answers, Code Quality, and Dropout on Stack Overflow

Elijah Zolduoarrati, Sherlock A. Licorish, Nigel Stanger
Department of Information Science
University of Otago
Dunedin, New Zealand
{elijah.zolduoarrati, sherlock.licorish, nigel.stanger]}@otago.ac.nz

**Abstract**: Previous studies that used data from Stack Overflow to develop predictive models often employed limited benchmarks of 3-5 models or adopted arbitrary selection methods. Despite being insightful, their limited scope suggests the need to benchmark more models to avoid overlooking untested algorithms. Our study evaluates 21 algorithms across three tasks: predicting the number of question a user is likely to answer, their code quality violations, and their dropout status. We employed normalisation, standardisation, as well as logarithmic and power transformations paired with Bayesian hyperparameter optimisation and genetic algorithms. CodeBERT, a pre-trained language model for both natural and programming languages, was fine-tuned to classify user dropout given their posts (questions and answers) and code snippets. We found Bagging ensemble models combined with standardisation achieved the highest $R^2$ value (0.821) in predicting users' answers. The Stochastic Gradient Descent regressor, followed by Bagging and Epsilon Support Vector Machine models, consistently demonstrated superior performance to other benchmarked algorithms in predicting users' code quality across multiple quality dimensions and languages. Extreme Gradient Boosting paired with log-transformation exhibited the highest $F_1$-score (0.825) in predicting users' dropout. CodeBERT was able to classify users' dropout with a final $F_1$-score of 0.809, validating the performance of Extreme Gradient Boosting that was solely based on numerical data. Overall, our benchmarking of 21 algorithms provides multiple insights. Researchers can leverage findings regarding the most suitable models for specific target variables, and practitioners can utilise the identified optimal hyperparameters to reduce the initial search space during their own hyperparameter tuning processes.

## 1 INTRODUCTION

The knowledge-intensive nature of software engineering (SE) presents developers with frequent technical challenges requiring knowledge they do not possess (Ye et al., 2017). To address these challenges, developers tend to leverage Stack Overflow[1], which hosts a vast collection of solutions to SE-related problems (Barua et al., 2014). Stack Overflow's distinctive model of peer-to-peer support and collaborative problem-solving has thus demonstrably shaped the SE landscape. This platform empowers developers to work together in overcoming programming challenges, ultimately driving advancements across the industry (Zahedi et al., 2020). The breadth of user-generated content on the website encompasses millions of contributors around the world posting and answering questions on a monthly basis (Ye et al., 2017). This wealth of information has led to numerous studies utilising Stack Overflow data for prediction tasks to model platform elements. Aspects such as question scores (Alharthi et al., 2016), gamification elements (Wang, Chen, et al., 2020), and duplicate posts (Ahasanuzzaman et al., 2016) have been subjected to modelling-related tasks. In spite of their insightful findings, many prior studies often limit their benchmarking to 3-5 models or select models arbitrarily without proper justification, such as Linear Regression (Matei et al., 2018; Morrison and Murphy-Hill, 2013) or between Random Forest and Neural Networks (Omondiagbe et al., 2019). This constitutes a limitation, and may overlook untested models that could potentially outperform those evaluated. Consequently, the research community may still be utilising models that are not necessarily optimal for the given task. Our study aims to bridge this gap and benchmark a broader catalogue of models, encompassing not only widely-used algorithms such as Random Forests or Logistic Regression (Adaji and Vassileva, 2015), but also models that are less popular in the literature, such as Nu-Support Vector Machine (SVM) and Theil-Sen estimators (Akritas et al., 1995; Schölkopf et al., 2000). Specifically, we establish the most optimal models when predicting *Answers*, *Code Quality*, and user *Dropout*. Our investigation culminates in leveraging a fine-tuned transformer-based model to compare the efficacy of numeric-based versus textual-based user *Dropout* prediction. Numeric-based prediction uses quantifiable user attributes (e.g., users' total questions and their popularity) to assess dropout propensity, while textual-based prediction analyses their posts (including both questions and answers) for specific phrases or word patterns indicative of dropout risk. To the best of our knowledge, our study is the first to benchmark more than 20 different algorithms – both for regression and classification – to predict Stack Overflow users' *Answers*, their *Code Quality*, and their *Dropout* status.

[1] https://stackoverflow.com

In this regard, our study contributes to both theory and practice. For theory, we bridge a methodological gap where studies of similar design frequently employ modelling techniques without applying a robust objective selection framework (see Section 2). Our study provides a catalogue of various modelling approaches, which we believe will benefit future researchers under this niche by providing guidance on selecting suitable models for specific tasks. For industry practitioners, our study provides readily-available models that can be leveraged to develop targeted interventions, potentially encouraging contributors to maintain interest. Practitioners can leverage our identified hyperparameters as a foundational point of reference (i.e., start point) prior to tuning their own models for similar tasks. We also contribute to the industry's understanding regarding the practicality of transformer-based models, whose use traditionally only revolves around question answering, text summarisation, and sentiment analysis (Zhao et al., 2023). We demonstrate the feasibility of transformer-based models for predicting user dropout within online communities, broadening its applicability.

It should be emphasised that our primary objective is not to enhance existing models, but rather to illustrate the performance variability across models in various contexts. Within the niche of benchmarking studies, our work extends similar studies where model selection is done arbitrarily or where limited comparisons are performed, neglecting feature engineering. Restricted benchmarking of a few models and/or arbitrary model selection limits our understanding of the performance of various modelling techniques, which we aim to demonstrate in the current work.

The remainder of this manuscript is organised as follows. Section 2 provides background for the study, reviews the literature, and introduces the research questions. Section 3 details the methodologies employed for data collection, processing, and analysis. Section 4 presents the study's findings in relation to each research question. Section 5 discusses and contextualises these findings, reflecting on their implications for theory and practice. Section 6 considers threats to validity, and Section 7 concludes with final remarks and suggestions for future research. A replication package is provided for those interested in further examining our research methodology and performing replication studies (Zolduoarrati et al., 2025a) (refer to Section 8).

## 2 BACKGROUND AND LITERATURE REVIEW

Stack Overflow serves as a valuable resource for SE practitioners seeking information. It leverages the collective knowledge of its user base, where practitioners are likely to encounter questions previously addressed by others facing similar challenges (Meldrum et al., 2017). This dynamic has fostered a wealth of research on Stack Overflow. For instance, Anderson et al. (2012) delved into how answers and votes influence the selection of the best answer within a given thread. Baltadzhieva and Chrupała (2015) explored the potential of leveraging question metadata, such as tags, title, and body length, to predict how well a question would be received on the platform. Alharthi et al. (2016) extended this by incorporating additional metadata to predict question scores, such as the presence of hyperlinks, the length of any included code snippets, and the time it took for the question to be answered.

Beyond studies investigating the multifaceted nature of *contribution*, others have also inspected the gamification side of Stack Overflow. Papoutsoglou et al. (2020) clustered users based on their badge levels (bronze, silver, gold) and reputation, along with their accepted and total answers, to identify how gamification relates to user activity. Wang et al. (2020) investigated the impact of Stack Overflow's badge system on user contributions. Their findings suggest that the badges primarily incentivise users to increase the quantity and frequency of their contributions, hinting at behaviours aimed at maximising badge acquisition rather than focussing on quality contributions. The focus on contribution quality extends to research on identifying redundant content. In one study, Ahasanuzzaman et al. (2016) proposed a classification technique to automatically detect duplicate questions, which is particularly relevant because Stack Overflow currently relies on its moderators to manually identify and flag such questions. In our work, we seek to predict three specific elements; users' answers (i.e., number of questions a user is likely to answer), code quality, and user dropout. The choice behind the selection of these three metrics as the core focus stems from several reasons. First, they fundamentally underpin the knowledge ecosystem's vitality, directly influencing content generation, information reliability, and community stability (Anderson et al., 2012; Srba and Bielikova, 2016). Answer production serves as the primary mechanism for knowledge transfer (Zolduoarrati et al., 2024), while code quality functions as the determinant of solution efficacy (Meldrum et al., 2020). User dropout indicate community health and sustainability (Srba and Bielikova, 2016). Second, these three metrics constitute core functional outputs rather than ancillary activities or gamification elements, which otherwise can be operationalised by edits or badges (Papoutsoglou et al., 2020; Zolduoarrati et al., 2024). For instance, while badges and reputation systems may drive engagement, they represent proxies for expertise rather than direct measures of platform value generation (Osborne et al., 2024).

Given the breadth of latent knowledge to be unearthed, numerous research endeavours have attempted to predict patterns stemming from user-generated content on Stack Overflow. Nevertheless, a recurring limitation in such studies is the reliance on narrow benchmarks or selection of models without proper benchmarking. This widespread issue introduces the threat where untested models may yield better results. In the next subsection, we examine related works that exemplify this gap. As our study seeks to harness transformer-based models in validating numerical-based models, we will also explore the prevalence of transformer-based models in recent literature.

## 2.1 Stack Overflow Contribution Prediction

Several recent works have focussed on predicting aspects related to Stack Overflow question-answering. Zhou et al. (2020) leveraged Logistic Regression to identify factors that influence the likelihood and speed of resolution for bountied questions. Ahasanuzzaman et al. (2018) proposed a sequential Logistic Regression model to classify posts that specifically deal with issues related to application programming interfaces (APIs). Tian et al. (2021) investigated the use of a Random Forest model to predict whether a given answer would be chosen as the best answer to a question thread. Matei et al. (2018) employed Linear Regression models to analyse how fluctuations in social factors on the platform might influence the calibre of knowledge generated by its users. Morrison and Murphy-Hill (2013) also harnessed Linear Regression and found an association between a user's biological age and reputation on Stack Overflow, suggesting that one positively influences the other.

Despite the valuable insights provided by these works, these studies suffer from a familiar methodological limitation: modelling techniques are often adopted in an ad-hoc manner where choices are made based solely on contextual considerations. Those that do undertake benchmarking efforts have been limited in scope, typically evaluating only a small number of models (e.g., 3 to 5). For example, Asaduzzaman et al. (2013) benchmarked Random Forest and J48 classifiers (two algorithms) to predict the duration of unanswered questions on Stack Overflow. Various question features, including votes, post length, readability, and the asker's score were all considered as predictors (Asaduzzaman et al., 2013). Yazdaninia et al. (2021) benchmarked six algorithms: Extreme Gradient Boosting, CART, Bayesian Ridge, Ridge, Lasso, and Gaussian Naive Bayes to predict whether a Stack Overflow question would receive an accepted answer or remain unresolved. Various features were subject to analysis, including semantic (e.g., number of paragraphs and links), question metadata (e.g., asking time), and asker-inherent features (e.g., badges and membership duration). Their findings revealed that Extreme Gradient Boosting was the top-performing model with an Area Under the ROC Curve (AUC) of 0.71 (Yazdaninia et al., 2021). Gama et al. (2023) employed Random Forest, Gradient Boosting, Extreme Gradient Boosting, Naive Bayes, and SVM (five algorithms) to classify technical debt within Stack Overflow discussions.

Based on the review above, it is apparent that previous research often neglects benchmarking a broad catalogue of models, leaving open the possibility that untested models may outperform tested ones. For example, one study by Omondiagbe et al. (2019) compared the performance between Random Forest and a Neural Network at predicting answers' acceptability, yet the relative performance of other ensemble algorithms, such as Gradient Boosted Trees or AdaBoost, remains unknown. Thus, the current body of knowledge lacks a robust understanding of which modelling techniques are most effective at predicting Stack Overflow contributions.

However, the term *contribution* itself is multifaceted, making it challenging to define a singular, universally-applicable metric to gauge such an abstract notion. As a result, prior studies have consistently recognised users' answers as a key form of contribution (Choetkiertikul et al., 2015; Mustafa et al., 2023; Zolduoarrati et al., 2024). For instance, Mustafa et al. (2023) harnessed users' total answers to operationalise voluntary knowledge contribution. Another example is by Wijekoon and Merunka (2022), where the average number of answers posted per user was one key component in quantifying voluntary knowledge contribution. Considering that answers are the core content of any CQA platform (Jang and Kim, 2024), it is unsurprising that answers are widely-regarded as a primary, if not the most critical, form of contribution. Thus, predicting users' answers is an important step in understanding one of the most crucial components of *contribution* within Stack Overflow. Accurate prediction of how many answers a user would give should then be the first focus of our study. The first RQ is formulated to guide our investigation: *What is the best modelling approach to predict users' answers?*

## 2.2 Stack Overflow Code Quality Prediction

Beyond question-answering, several studies have explored code snippets to model Stack Overflow contributions. Code snippets enhance the presentation quality and provide contextual details (Yazdaninia et al., 2021). For instance, within questions, code snippets may shed light on less explicit background information (e.g., libraries used or the expected code behaviour). Duijn et al. (2015) benchmarked Random Forest, Decision Tree, and Logistic Regression to predict post quality. Their findings revealed that the code-to-text ratio is the most significant factor compared to other semantic aspects like readability and clarity (Duijn et al., 2015). Correa and Sureka (2013) benchmarked SVM, Naive Bayes, Logistic Regression, and Stochastic Gradient Boosted Trees (SGBT) to predict whether a Stack Overflow question would be closed. Their analysis revealed that the length of code snippets in questions is a significant factor, with closed questions tending to have shorter code snippets compared to non-closed questions (Correa and Sureka, 2013). Ponzanelli et al. (2014) used a Linear Quality Function combined with min-max normalisation to classify questions as very bad, bad, good, or very good. A key feature considered in this classification was the percentage of lines of code declared within code snippets throughout the question text (Ponzanelli et al., 2014). Beyond these studies, there is a notable lack of research specifically focussed on predicting code quality. This may be due to the inherent challenge of defining code quality (Meldrum et al., 2020), as the quality of code can vary significantly across different programming languages (Zolduoarrati et al., 2025b).

Given the limited research on code quality prediction, our study aims to address this gap by assessing which model is the most optimal at predicting code quality in Stack Overflow posts. Building upon previous work (Zolduoarrati et al., 2025b), we define *Code Quality* to encompass four dimensions: reliability, readability, performance, and security. These quality dimensions are studied across five languages: SQL, JavaScript, Python, Ruby, and Java. These five languages are widely used in the US (Zolduoarrati et al., 2025b), justifying the need to understand quality implications for such snippets' reuse. To the best of our knowledge, our study is the first to predict code quality on Stack Overflow, incorporating all four dimensions across five languages. To address this opportunity, we formulate the second RQ: *What is the best modelling approach to predict users' code quality?*

## 2.3 Stack Overflow Dropout Prediction

Another significant area of research utilising Stack Overflow data is the prediction of user churn. Previous studies have used the terms "dropout" and "churn" interchangeably to refer to a decline in user activity over time (Adaji and Vassileva, 2015). Adaji and Vassileva (2015) benchmarked Logistic Regression, Neural Networks, SVM, and Random Forest to predict dropout among experts on Stack Overflow. These experts are characterised by high reputations, privileges, and the number of "best answer" awards they had received. Adaji and Vassileva (2015) found that Random Forests outperformed the other models, achieving an AUC of 0.82. Pudipeddi et al. (2014) explored the possibility of identifying early indicators of user dropout based on their initial posts or days of activity. The authors benchmarked Decision Tree, Logistic Regression, as well as SVM with linear and radial basis function kernels to classify users and found that Decision Tree outperformed the other three models (Pudipeddi et al., 2014). Mahbub et al. (2021) employed an SVM to classify whether a user would continue participating based on their answers, comments, downvotes, upvotes, and average post polarity.

However, similar to the limitations observed in studies on question-answering prediction (see Section 2.1), many of the aforementioned studies either chose models arbitrarily, or employed limited benchmarks of 3-5 models, again increasing the risk of overlooking untested models that may outperform those evaluated. Predicting users' sustained participation is no less important than predicting their answers or code quality, because accurately mitigating churn is essential to sustain the longevity of question-and-answer platforms (Pudipeddi et al., 2014). Moreover, Stack Overflow specifically continues to face challenges related to user dropout (Srba and Bielikova, 2016). For example, the emergence of generative pre-trained transformer (GPT) chatbots like ChatGPT threatens the platform's relevance, as some users prefer the comprehensive and well-articulated responses provided by these AI tools regardless of whether they are correct (Kabir et al., 2024). Consequently, it is worthwhile to identify the most effective model for predicting user dropout. The third RQ is formulated accordingly: *What is the best modelling approach to predict users' dropout?*

Following the identification of the most effective modelling approaches for the three RQs, we intend to export the models developed so that they will be readily available for use by the community, reducing the need to conduct data collection, processing, and subsequent training regimes.

## 2.4 Transformer-Based Models for Dropout Prediction

Transformer-based models have revolutionized the field of natural language processing (NLP) since their introduction by Vaswani et al. (2017). Transformer-based models represent a significant leap forward in the field of NLP, where previously, NLP-related tasks relied on conventional statistical approaches like n-gram models or Recurrent Neural Networks (RNNs). However, these earlier models struggled to capture long-range dependencies and contextual nuances within language. N-gram models, for instance, are limited by their consideration of a fixed sequence of words, hindering their ability to grasp complex relationships between words that might be distant in a sentence (Fink, 2014). Transformer-based models employ a self-attention mechanism that enables parallel processing of sequential data while capturing long-range dependencies between tokens, effectively addressing the computational limitations of previous state-of-the-art implementations (Vaswani et al., 2017). The architecture of transformers allows it to dynamically focus on relevant parts of the input regardless of their position in the sequence (Singh and Raman, 2024). Consequently they have accrued significant interest in recent years due to their versatility across diverse applications, including sentiment analysis, topic classification, and intent recognition (Bashiri and Naderi, 2024; Benayas et al., 2021)

Bidirectional Encoder Representations from Transformers (BERT) is one of the earliest transformers, which leverages a self-attention mechanism that empowers the model to analyse the relative importance of individual words within a sentence. Such an architecture enables it to capture intricate, implicit context within textual data (Raiaan et al., 2024). BERT's versatility is attributed to its fine-tuning capability, which involves adding an output layer specifically tailored to the desired task. This led BERT to excel in domains such as question-answering and language inference (Raiaan et al., 2024). Following the widespread success of BERT, the scientific community has developed numerous successor models, including RoBERTa and DistilBERT (Adoma et al., 2020), built upon BERT's foundation and optimised to address a broad range of tasks across diverse domains. However, despite the success of BERT and its successor models in sentiment analysis and quality prediction within Stack Overflow data (Biswas et al., 2020), their application is typically limited to purely natural language inputs. To bridge the gap between natural and programming languages, models like CodeBERT have been proposed (Feng et al., 2020). CodeBERT is a state-of-the-art model trained on both natural and programming languages, allowing it to learn word representations that effectively capture the nuances of both domains. The model has demonstrated good effect on SE-related tasks like natural language code search and code documentation generation (Feng et al., 2020). CodeBERT leverages a bidirectional architecture to capture contextual information surrounding each code and natural language token, enabling learning of long-range dependencies.

Previous research has successfully used CodeBERT fine-tuned on Stack Overflow data. He et al. (2022) demonstrated CodeBERT's superior performance compared to other BERT variants in recommending tags for Stack Overflow posts based on their titles and code snippets. Luong et al. (2021) employed CodeBERT to extract the semantic meaning of each paragraph/code snippet pair, enabling the identification of API-related discussions on the platform. Zhou et al. (2021) used CodeBERT to identify the most semantically similar code snippet to a given query, within a collection of candidate code snippets extracted from Python-related posts. However, to the best of our knowledge, previous works have not used CodeBERT to predict user dropout based on their post history, taking into account both their textual contributions (i.e., communicated paragraphs) and code snippets. Our study aims to address this gap, and thus we formulate the fourth RQ: *How does a transformer-based model performs when compared to the chosen numerical model in predicting user dropout?*

The next section details the methods employed in our study.

# 3 METHODS

## 3.1 Data Collection

To ensure focussed yet fruitful downstream analyses, we restrict our current investigation to users residing within the United States (US) for several key reasons. First, an earlier study (Zolduoarrati et al., 2024) revealed a consistent propagation of trends from global to city levels and developed a high-quality dataset suitable for reuse. Second, the US is the world's largest English-speaking nation (Yeh, 2022), rendering it a suitable environment to study social computing, where English is also the *lingua franca* (Lutz, 2009). The US also consistently ranks at the forefront of technological innovation (Broughel and Thierer, 2019), contributing to a diverse developer pool that acts as a microcosm for global SE-related rigours. Finally, the US boasts one of the largest user bases on Stack Overflow compared to other countries, making it a representative platform to portray the dynamics of community question-answering (CQA) settings (Zolduoarrati et al., 2024). Within this defined scope, we also build on prior works (Zolduoarrati et al., 2025b, 2024, 2025c).

A user-level dataset was constructed based such prior works, consisting of information specific to individual users on the platform; platform variables (e.g., total *Questions* and *Badges*), inherent user factors (e.g., *User Popularity* and *Profile Completion Rate*), and semantic aspects (e.g., *Post Polarity* and *Readability*). This dataset was used to benchmark a range of modelling techniques to predict users' *Answers*, the *Code Quality* of their snippets, and their *Dropout* status. Table 1 provides definition of each variable used in the analysis. Variables designated as target variables (labels) in the subsequent RQs are marked with the superscript letter *a*. Specifically, *Code Quality*-related target variables marked with the superscript letter *b* refer to a set of metrics used in prior work (Zolduoarrati et al., 2025b) to evaluate the quality of code written in different programming languages. These metrics encompass four key dimensions (Reliability, Readability, Performance, and Security) and are calculated for five specific languages (SQL, JavaScript, Python, Ruby, Java), so there are 20 such target variables. For instance, one target variable is *Java Reliability Violation Density*, which measures the frequency of reliability issues encountered within Java code. Similarly, the *Python Performance Violation Density* target variable assesses the prevalence of performance inefficiencies in Python snippets. Unmarked variables were otherwise used as predictor variables (features).

Given that this study utilises two distinct datasets – one containing user-level metrics to predict users' *Answers* (RQ1), their *Code Quality* (RQ2), and *Dropout* (RQ3) and the other encompassing textual data from user posts for transformer-based evaluation (RQ4) – we categorise the scope as either numeric-based or textual-based predictions (Ahasanuzzaman et al., 2018). The former corresponds to both regression (RQ1 and RQ2) and classification tasks (RQ3) whereas the latter focusses on textual-based prediction using transformer-based model (RQ4), which is a classification task. This distinction arises from the variables' scales: users' *Answers* and *Code Quality* are measured on at least an interval scale and are therefore non-categorical, whereas user *Dropout* is measured on a nominal scale (i.e., binary) and is thus categorical. Table 2 provides a detailed breakdown of the tasks and dataset associated with each RQ. After this initial data acquisition phase, we proceeded with data processing which is outlined in the next section.

Table 1. Origin, definition, and computation details of study variables

| Study Origin | Variable Name | Definition |
|---|---|---|
| *Harmonising Contributions: Exploring Diversity in Software Engineering through CQA Mining on Stack Overflow* (Zolduoarrati et al., 2024) | *Gender* | User's inferred binary gender, using Genderize[2] (Zolduoarrati and Licorish, 2021) |
| | *ProfileLength* | Number of words users wrote about themselves in their profile. |
| | *YearlyDurationUsage* | Number of years between a user's initial registration to their last access date. |
| | *UpVotes* | Total upvotes cast by a user to indicate useful posts. |
| | *DownVotes* | Total downvotes cast by a user to indicate posts of minimal value. |
| | *Views* | Total number of times a user's profile has been viewed. |
| | *Reputation* | Total number of scores from a user's posts. |
| | *Questions* | Total questions asked by a user. |
| | *Answers* [a] | Total answers provided by a user. |
| | *Comments* | Total comments provided by a user. |
| | *Edits* | Total edits a user made to existing posts/comments. |
| | *Badges* | Total badges earned by a user, regardless of tiers. |
| *Stack Overflow's Hidden Nuances: How Does Zip Code Define User Contribution?* (Zolduoarrati et al., 2025c) | *User Development Index* | Sum of users' *Questions* and *Answers*. Depicts the intensity of question-asking and answering practices. |
| | *User Management Index* | Sum of users' *UpVotes* and *DownVotes*. Depicts the efforts poured into content management. |
| | *User Contribution Frequency* | Daily aggregate of users' *Questions* and *Answers*. |
| | *User Disengagement Rate* | Length of time between a user's last visit and the current date, flagging them as disengaged if it surpasses 24 months. |
| | *User Dropout* [a] | Binary variable indicating whether a user has stopped actively participating after 60 days from their initial question or answer. |
| | *Post Readability* | The cognitive effort exerted to comprehend a post. A score of 100 signifies a minimal cognitive load, and 0 otherwise. |
| | *Post Attention to Detail* | The degree of typographical errors within posts. |
| | *Post Polarity* | The extent of affirmation (positive emotion) or negation (negative) within posts. |
| | *User Profile Completion Rate* | The extent to which users disclose personal information within their profiles. |
| | *Code Length* | A measure to depict the complexity and length of problems being discussed. |
| | *User Popularity Index* | A measure to reflect users' popularity and renown within the community. |
| *Does Location Influence Coding Practices? A* | *Reliability Violation Density* [b] | Users' code reliability violations, averaged over their total logical lines of code (LLOC). |

[2] https://genderize.io

| Study Origin | Variable Name | Definition |
|---|---|---|
| *Cross-Regional Study on Stack Overflow Code Quality* (Zolduoarrati et al., 2025b) | *Readability Violation Density* [b] | Users' code readability violations, averaged over their total lines of code (LOC). |
| | *Performance Violation Density* [b] | Users' code performance violations, averaged over their total LLOC. |
| | *Security Violation Density* [b] | Users' code security violations, averaged over their total LLOC. |

[a] Target variables (labels).
[b] Code quality-related target variables, analysed with respect to all five languages.

Table 2. Tasks and datasets for each RQ

| RQ# | Prediction Scope | Dataset Employed | Task | Target Variable |
|---|---|---|---|---|
| RQ1 | Numeric-based | Users dataset | Regression | Users' total *Answers* |
| RQ2 | | | | Users' *Code Quality* violation density |
| RQ3 | | | Classification | Users' *Dropout* |
| RQ4 | Textual-based | Users' posts dataset | Classification | |

### 3.2 Data Processing

Our data processing pipeline separates numeric and textual features. For numeric features to answer RQ1 to RQ3, we prioritised handling of missing values as many models are sensitive to missing data, and improper imputation may lead to inaccurate results. To ensure scientific rigour, the approach to handling missing values depends on the underlying meaning of each variable in Table 1. For instance, missing values in *Comment Polarity* signify a lack of comments, not neutral sentiment. Consequently, imputation with zeros would be misleading as it would imply the latter. Therefore, a one-size-fits-all approach to missing value handling would be inappropriate. Instead, the most suitable methods were determined for each unique variable, weighing their specific meaning, implications, and inherent characteristics. Not all variables required missing value imputation as some variables had no missing data (e.g., *YearlyDurationUsage*). Therefore, imputation efforts were targeted only at those with missing data, depicted in Table 3. Explanations for each strategy are detailed in our replication package (Zolduoarrati et al., 2025a)[3].

As seen in Table 3, for *User Contribution Frequency*, imputation was done using K-Nearest Neighbours. A k-value (i.e., number. of neighbours) was selected to be 5 based on empirical trial-and-error results (Beretta and Santaniello, 2016). The Euclidean distance metric was selected based on its demonstrated efficacy when combined with this imputation methodology (Beretta and Santaniello, 2016). Finally, inverse distance weighting was used as it ensures that closer neighbours have proportionally greater influence on imputed values, preserving local pattern recognition (Bhowmick and Saha, 2023). For variables that were imputed using Expectation Maximization, tolerance threshold was set to 1e-6 for good precision (Zhou et al., 2017). Maximum iterations were increased to 100, ensuring adequate convergence opportunity compared to lower thresholds of 25 or 50 iterations. Random initialisation of values was employed to optimise both outcome quality and convergence speed (Barazandeh and Razaviyayn, 2018). We recognise that our imputation strategies may not be perfect, which will be discussed in Section 6.

Table 3. Missing value imputation strategies

| Study Origin | Variable Name | Implications | Strategy | Rationale |
|---|---|---|---|---|
| *Harmonising Contributions: Exploring Diversity in Software Engineering through CQA Mining on Stack Overflow* (Zolduoarrati et al., 2024) | *ProfileLength* | Lack of measurement | Zero imputation | Null values correspond directly to an absence of the associated action. For instance, a null value in *Comments* indicates the user authored zero comments. |
| | *UpVotes* | | | |
| | *DownVotes* | | | |
| | *Views* | | | |
| | *Reputation* | | | |
| | *Questions* | | | |
| | *Answers* [a] | | | |
| | *Comments* | | | |
| | *Edits* | | | |
| | *Badges* | | | |
| *Stack Overflow's Hidden Nuances: How Does Zip Code Define User Contribution?* | *User Contribution Frequency* | Unavailable data | K-Nearest Neighbours imputation | Users with similar profiles may share similar posting frequencies. |
| | *Post Readability* | Incomplete operationalisation | | |
| | *Post Attention to Detail* | | | |

[3] Replication package » Preprocessing and Model Selection » Missing Value Handling.docx

| Study Origin | Variable Name | Implications | Strategy | Rationale |
|---|---|---|---|---|
| (Zolduoarrati et al., 2025c) | *About Me Polarity* | | Expectation Maximization (EM) imputation | Users with similar profiles may share similar behavioural tendencies. |
| | *Comment Polarity* | | | |
| | *Question Polarity* | | | |
| | *Answer Polarity* | | | |
| | *User Popularity Index* | | | |
| | *Code Length* | Lack of measurement | Zero imputation | Null values indicate no code snippets in their user-generated content, which is essentially zero code characters. |
| *Does Location Influence Coding Practices? A Cross-Regional Study on Stack Overflow Code Quality* (Zolduoarrati et al., 2025b) | *Violation Density* [b] | Lack of measurement | Zero imputation | Null values indicate an absence of code quality issues across four assessed dimensions within the users' answers, which is essentially zero code violations. |

[a] Target variables (labels).
[b] Code quality-related target variables, analysed with respect to all five languages and four dimensions.

For RQ4, which explores the potential of transformer-based models, we utilise users' posts (i.e., questions and answers) as the primary data source, ensuring alignment with the methodology employed in previous works (Zolduoarrati et al., 2025b, 2025c). RQ4 principally aims to validate the numerical-based classification model obtained in RQ3. We only validated RQ3 because transformer-based models were predominantly used for text classification tasks in similar studies (He et al., 2024), while RQ1 and RQ2 pertain to regression tasks. We acknowledge that this means that RQ1 and RQ2 remain unvalidated, a limitation that will be discussed in Section 6. Afterwards, we parsed HTML tags using the BeautifulSoup4 Python library – extracting both textual content within paragraph tags (situated between <p> tags) and code snippets enclosed in preformatted code tags (<pre><code> tags) (Meldrum et al., 2020), thereby guaranteeing that no relevant information from user posts was overlooked. In total, there were 9,467,264 snippets across 1,601,561 questions and 3,711,516 answers. Following data processing, we adopted methods for the specific prediction scope (see Table 1) for each RQ. Our user data (RQ1 – RQ3) encompasses numeric features, and underwent preprocessing techniques suited only for numerical data. Conversely, RQ4, which leverages textual data from user posts, necessitated text-specific preprocessing methods such as stop words removal and tokenisation directly using the built-in CodeBERT tokeniser (explained in Section 3.5). The next section sheds light on our experimental design based on good practices in literature.

### 3.3 Experimental Design

To ensure scientific rigour, we first conduct a brief literature review to understand existing best practices in the domain, encompassing both numeric and textual-based scope. The insights gained informed our research methodology, as outlined in Table 4. Measures marked by the superscript letter *a* are only applicable to numeric-based prediction, where those marked with the superscript letter *b* are only applicable to textual-based prediction. Unmarked measures are applicable to both. Figure 1 depicts our experimental design.

Table 4. Existing best practices

| Task | Measure | Description | Merit | Example(s) |
|---|---|---|---|---|
| Preprocessing | Data cleaning | Systematic identification and correction of errors (Van den Broeck and Fadnes, 2013). | Enhances the reliability of the subsequent analyses and conclusions (Van den Broeck and Fadnes, 2013). | Missing value imputation, comment removal. |
| | Feature engineering | Selection and transformation of raw data into more informative features (Khurana et al., 2018). | Enhances a machine learning (ML) model's predictive accuracy for a given task | Normalisation and standardisation, collinearity detection, tokenisation. |

[4] https://pypi.org/project/beautifulsoup4

| Task | Measure | Description | Merit | Example(s) |
| --- | --- | --- | --- | --- |
| | | | (Khurana et al., 2018) | |
| | Data splitting | Splitting the entire dataset with a predetermined ratio, where the majority of data is used for training, and the rest for testing (Pargent et al., 2023). | Avoids information leak and provides robust assessment on well the model performs on unseen data (Raschka, 2018). | 80/20 train/test splitting. |
| Modelling | Baseline modelling [a] | Meta-estimator that does not learn any patterns from the data (Pargent et al., 2023). | Allows performance benchmarking against more complex ML models (Pargent et al., 2023). | Regressors that predict the mean target value, classifiers that predict the majority class label. |
| | Transfer learning [b] | Fine-tunes models that have been pre-trained on a large, general dataset as a starting point. | Adapts pre-existing knowledge to the new task, thereby reducing the training time required to obtain good generalisation of the data (Cook et al., 2013). | Usage of pre-trained transformer-based models such as BERT variants. |
| | Hyperparameter tuning | Finds the optimal configuration settings (i.e., hyperparameters) of ML models (Raji et al., 2022). | Improves the performance of ML models (Raji et al., 2022). | Bayesian optimisation, genetic algorithm. |
| | Resampling | Iteratively splitting different partitions of the data to use as test/training sets (Pargent et al., 2023). | Mitigates bias and reduces variance, leading to a more generalisable model (Pargent et al., 2023). | k-fold cross-validation, oversampling for imbalanced classes. |
| | Training callbacks | Actions to automatically perform at various training stages. | Customises the behaviour of training such that efficiency is improved and enables mining latent insights. | Reducing learning rate on loss plateau, stop training when the best accuracy scores are obtained. |
| Reporting | Multifaceted evaluation | Reporting of multiple evaluation metrics instead of relying on a single one (Heyburn et al., 2018). | Guards against misleading conclusions and ensures a more robust assessment of the models' generalisability (Heyburn et al., 2018) | RMSE and $R^2$ for regression, accuracy for classification. |
| | Best-fit hyperparameters | Reporting the complete hyperparameters that result in the most optimal model. | Ensures reproducibility and enables comparability across other models (Lynnerup et al., 2020). | (In Random Forests) number of estimators, maximum tree depth. |

[a] Applicable only for numeric-based prediction.
[b] Applicable only for textual-based prediction.

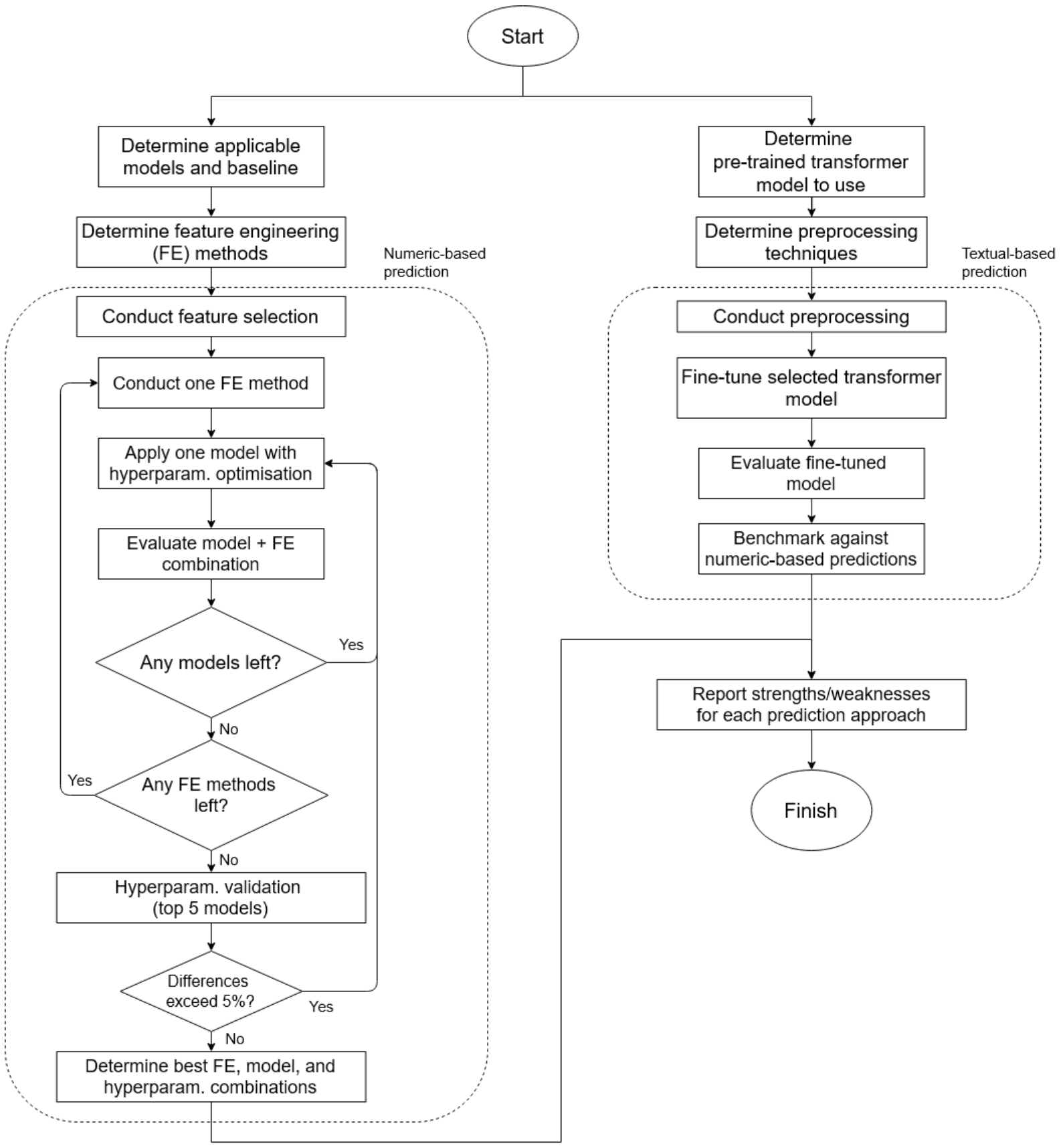

Figure 1. Study design

For all tasks, we partitioned the data into training and testing sets using an 80/20 split. We used separate validation sets as all implemented algorithms incorporate k-fold cross-validation for model assessment, which increases generalisability (Omondiagbe et al., 2019). The value of k (number of folds) is set to 4 (Omondiagbe et al., 2019). Next, to inform the creation of classes for subsequent classification tasks, we determined from prior work that the US user base on Stack Overflow exhibits a significant disparity in contribution levels (Zolduoarrati et al., 2025c). Among the 760,809 total users, only 16,813 (2.21%) qualify as active contributors who have not dropped out, consistently providing questions or answers. The vast majority of 743,996 users (97.79%) have instead dropped out. Thus, we assigned the active contributors as the positive class and the dropped-out users as the negative, aligning with best practices where the minority class is usually regarded as positive (Sain and Purnami, 2015). Finally, to mitigate biases introduced by class imbalance when answering RQ3, we adopted the Synthetic Minority Oversampling Technique (SMOTE), shown to yield more effective results compared to other oversampling techniques, such as Random Oversampling (ROS) or Adaptive Synthetic Sampling (ADASYN) (Omondiagbe et al., 2019). SMOTE was applied with a sampling ratio of 0.5 (i.e., 2:4 ratio of minority to majority instances), aiming to strike optimal balance between providing sufficient representation of non-dropout users while avoiding excessive synthetic data generation that could lead to overfitting (Dankar and Ibrahim, 2021). Standard SMOTE was applied – as opposed to specialised variants like SMOTE-NC – given that our feature spaces are primarily continuous, not discrete nor categorical (Alonso et al., 2025). Implementations were done using the third-party package imbalanced-learn[5], with this package having shown good configurability and scalability previously (Wongvorachan et al., 2023).

### 3.4 Numeric-Based Predictions

For RQ1, RQ2, and RQ3, we employed a progressive feature selection strategy, incrementally expanding the predictor set (i.e., features) for each RQ. First, to predict user's *Answers* in RQ1, we leveraged all platform variables and contribution metrics except *Code Quality* (addressed in RQ2) and *Dropout* (addressed in RQ3). Building on this foundation, RQ2 incorporates users' *Answers* – which was the target variable in RQ1 – as an additional predictor alongside the existing features from RQ1. Finally, for predicting user *Dropout* in RQ3, we introduced *Code Quality*-related variables as a new feature set, complementing the features established in RQ1 and RQ2. We argue that this cumulative approach ensures maximal utilisation of available information to develop robust prediction models for each RQ, facilitating a stepwise progression from one RQ to the next. The target variables of RQ2 and RQ3 can theoretically serve as predictors for RQ1 and RQ2, respectively. However, we believe that such an approach would be redundant as the target variable of each RQ will subsequently be utilised as a predictor in the following RQ. For instance, users' *Code Quality* and their *Dropout* could conceivably impact users' *Answers* – yet including them as predictors would be unnecessary as the study investigated how users' *Answers* predict *Code Quality* and *Dropout* in the later models anyway (RQ2 and RQ3). Nevertheless, both may serve as good predictors of users' *Answers*, and our highest-performing model for RQ1 may not exhibit generalisability across all scenarios as it overlooks the influence of *Code Quality* and user *Dropout* (see Section 6).

[5] https://imbalanced-learn.org

As depicted in Figure 1, a set of feature engineering (FE) and comparative modelling techniques were applied independently to RQ1, RQ2, and RQ3. This means each of the three RQs underwent its own distinct FE process and comparative modelling phase. A rigorous evaluation strategy was then conducted to identify which FE/modelling combinations yielded the most accurate and robust predictions to answer each RQ. Specifically for *Code Quality* violations (RQ2), we conducted a total of 20 distinct sets of model comparisons, given we have 5 languages across 4 quality dimensions. The rationale is that model generalisability and performance may not carry over to different dimensions and languages. For example, a model that excels at predicting SQL reliability violations may not perform as effectively when predicting JavaScript security violations.

### *3.4.1 Baseline and Comparative Modelling*

Given the extensive range of well-established algorithms readily available within Python packages (e.g., *pytorch*[6] and *statsmodels*[7]), we restricted our models to implementations offered by Python-based libraries. This ensured a standardised analysis, avoiding unfairness that might arise from incorporating algorithms from other languages with varying implementation characteristics. Afterwards, we reviewed existing literature to identify a broad range of applicable models, assessing the relevance and merit of each model for inclusion in our study. This encompassed both shallow and deep learning techniques, as our primary goal was to achieve an exhaustive evaluation and catalogue each model's strengths and weaknesses, leaving no viable modelling approach unexplored (Omondiagbe et al., 2019). For instance, while deep Neural Networks may offer superior performance, they often come at the cost of increased computational demands and resource requirements for training (e.g., Nvidia GPUs), which some researchers and practitioners may not have access to.

For regression tasks (RQ1 and RQ2), our selection process resulted in a shortlist of 18 regression algorithms for evaluation, whereas to conduct classification (RQ3), we evaluated 12 models (Table 5). In aggregate, our study comprised 21 distinct models. It is worth noting that applying all 21 models across all three RQs is methodologically infeasible. From these 21 models, 9 models can only apply to regression tasks (e.g., ordinary least squares), 3 models can only be fitted to classification problems (e.g., Ridge Classifier), and 9 other models can accommodate both regression and classification requirements (e.g., Decision Trees and Random Forests). To distinguish one from the other, models only applicable for regression tasks are marked with the superscript letter *a* in Table 5, while those only applicable for classification are marked the superscript letter *b*. Models marked with the superscript letter *c* are applicable for both cases. Strategically deploying models within their intended analytical domains – regression or classification – ensures methodological integrity by allowing each algorithm to operate within its optimal functional constraints, rather than imposing unfair tasks to which they were not designed for.

Next, some algorithms were excluded from the evaluation following our brief literature review, as they were superseded by more contemporary and objectively superior alternatives. For example, ElasticNet incorporates aspects of Lasso (L1) and Ridge (L2) regression, which effectively remedies the latter two's weaknesses (Zou and Hastie, 2005), and so both Lasso and Ridge were not considered. Another example is LARS which was superseded by Lasso Least Angle Regression (LassoLARS), which integrates Lasso's feature selection capabilities with LARS' computational efficiency, resulting in superior performance (Singh et al., 2021). Similarly, Gradient Boosting is a powerful technique, yet it suffers from computational limitations. It was superseded by Extreme Gradient Boosting and therefore only the latter was included in our benchmarks (Malallah and Abdulrazzaq, 2023). Moreover, some algorithms do not fit the measurement scale of our data.

[6] https://pytorch.org

[7] https://www.statsmodels.org

For example, Poisson regressors were originally designed to capture discrete data (e.g., counts) for their target variable (Ghazwani and Begum, 2023) instead of continuous data, which does not align with our study design where target variables are of ratio scale. Another example was Quantile regressors, designed to estimate conditional quantile functions (Xu, 2023). This diverges from our study's interest in predicting actual real values (e.g., users' *Answers* and *Code Quality* violations) and thus Quantile regressors were excluded from benchmarking. The full breadth of inclusion/exclusion rationale may be seen in our replication package (Zolduoarrati et al., 2025a)[8]. Despite the lengths taken to select models for our study, it is possible that some excluded models may yield better results, which we will discuss in Section 6.

With regard to the Neural Network to use for both regression and classification tasks, we first conducted a brief evaluation of several established architectures documented within the relevant literature, alongside our own custom architecture that incorporates best practices and draws inspiration from studies of similar nature (Omondiagbe et al., 2019; Salminen et al., 2020; Shen and Lee, 2016). This revealed that our custom architecture presented superior performance compared to replicating established architectures in their entirety, aligning with the well-documented principle that there is no universally optimal Neural Network architecture (Pham et al., 2020). Instead, customisation tailored to the specific characteristics of the problem at hand can often lead to superior outcomes. Our exploration is further detailed in Table 6, while benchmarking results for each architecture are seen in each RQ's corresponding subsections in Section 4.

Our custom implementation incorporates several strategies to enhance generalisability. First, we employ dropout layers with a rate of 0.2 after each hidden layer, which has been demonstrated to effectively prevent overfitting (Salminen et al., 2020). We exclude Nesterov momentum from the SGD implementation to avoid introducing redundancy (Shi, 2021). Finally, to optimise the training regime, we implemented exponential decay to adjust the learning rate – a technique proven more effective than fixed reductions (Lewkowycz, 2021). We terminate training after 10 epochs if no improvement is observed on the validation set, as 10 epochs are generally adequate to evaluate model progress and reach accuracy saturation (Nugraha et al., 2023).

To fairly evaluate shallow learning algorithms, we adopted literature recommendations to use a dummy baseline (Radabaugh et al., 2020). This simple model, often predicting the majority class in classification or a constant value in regression, enables us to scientifically assess the proposed models' effectiveness and track improvements during development, and provides a point of reference to objectively assess the efficacy of the proposed methods (Whigham et al., 2015). Comparing the predictive capabilities of the new method against a known, deterministic baseline may thus enable researchers to determine if it offers a statistically significant improvement in prediction accuracy. We leveraged the *DummyRegressor*[9] (for RQ1 and RQ2) and *DummyClassifier*[10] (for RQ3) from the *scikit-learn*[11] Python package to establish baseline models. These baseline models will be used as a reference point for regression and classification models that do not employ Neural Network architectures. On the other hand, given the unfairness of benchmarking complex Neural Networks against shallow baseline estimators, we constructed an identical network architecture but froze its weights during the prediction phase. This process essentially transforms the Neural Network into a baseline model that prevents the network from adapting to new data during prediction (Eberhard and Zesch, 2021), thereby establishing a benchmark that reflects the capabilities of the chosen architecture without the influence of learned patterns. All baseline models provide deterministic outputs – regressor baselines always predicts the mean of the training data, whereas those for classification always predicts the most frequent class label, as recommended by literature (Whigham et al., 2015).

---

[8] Replication package » Preprocessing and Model Selection » Included-Excluded Models.docx

[9] https://scikit-learn.org/stable/modules/generated/sklearn.dummy.DummyRegressor.html

[10] https://scikit-learn.org/stable/modules/generated/sklearn.dummy.DummyClassifier.html

[11] https://scikit-learn.org

Table 5. Shortlisted numeric-based models

| Umbrella Method | Method Name | Description | Merit for Inclusion |
|---|---|---|---|
| Linear models | Ordinary Least Squares (OLS) [a] | Foundational linear regression model. | Serves as the fundamental regression model due to its simplicity, allowing for transparent assessment and straightforward statistical inference (Kashki et al., 2021). |
| | Logistic Regression [b] | Probability predictor of an instance belonging to a given class. | Represents the canonical probabilistic framework for binary classification problems with well-established theoretical foundations, making it an essential benchmark against more complex classification models (Khan et al., 1999). |
| | ElasticNet [a] | OLS with combined L1 and L2 priors. | Incorporates both L1 and L2 regularisation methods to simultaneously perform variable selection and coefficient shrinkage, offering an optimal balance between feature elimination and multicollinearity management (Kashki et al., 2021). |
| | Ridge Classifier [b] | Converts the target values into {-1, 1}, then feeding a Ridge regression model. | Implements L2 regularisation within a classification context to reduce model variance and prevent overfitting due to multicollinearity, whilst also maintaining computational efficiency (van Laarhoven, 2017). |
| | LassoLARS [a] | Lasso model fit with least angle regression. | Combines Lasso's feature selection capability with LARS' computational efficiency, providing better predictive performance in high-dimensional spaces without exhausting computational resource (Benoit et al., 2021). |
| | Stochastic Gradient Descent (SGD) [c] | Linear model with stochastic gradient descent. | Demonstrates superior performance over traditional methods by incrementally optimising the objective function with randomly selected subsets of training data, making it particularly effective for large-scale learning problems (Abida et al., 2024). |
| Bayesian models | Bayesian [a] | Ridge with prior spherical Gaussian distribution. | Employs a Bayesian framework with spherical Gaussian priors to perform principled L2 regularisation without assuming data normality, allowing for uncertainty quantification (Bedoui and Lazar, 2020). |
| | Automatic Relevance Determination (ARD) [a] | Ridge with prior elliptical Gaussian distribution. | Utilises a hierarchical Bayesian approach with elliptical Gaussian priors to automatically identify and prune irrelevant features, providing a probabilistic alternative to traditional feature selection methods like VIF (Wipf and Nagarajan, 2007). |
| Outlier-robust models | Huber [a] | Linear model with L2 regularisation, robust to outliers. | Implements a robust loss function that reduces the influence of outliers without excluding them entirely, preserving latent information, maintaining regression performance in the presence of anomalous data points (Ge et al., 2019). |
| | Theil-Sen [a] | Linear model for noisy data, taking the median slope among all pairs of data points. | Provides a non-parametric approach that calculates median slopes between all possible data point pairs, offering resilience against both outliers and heteroscedasticity without requiring distributional assumptions (Helsel et al., 2020). |
| Support vector machines | Epsilon SVM [a] | SVM to achieve a specific tolerance level ($\mathcal{E}$) for training errors. | Employs a specific tolerance threshold (ε) for training errors, balancing predictive accuracy with computational efficiency (Majrashi et al., 2024). |
| | Nu SVM [a] | SVM to control the fraction of training errors and the fraction of support vectors. | Offers an alternative SVM parameterisation that directly controls the proportion of support vectors and training errors, which achieved better accuracy compared to other SVM variants (Majrashi et al., 2024). |
| | Linear SVM [c] | SVM to find a perfect linear fit. | Provides an optimal hyperplane for linearly-separable data while maintaining good generalisation capability, especially in scenarios where non-linear relationships are not predominant (Juang and Hsieh, 2012). |
| | C-SVM [b] | Linear SVM that balances margin and error via a cost parameter $C$. | Superior generalisation ability given non-linear and high-dimensional datasets (Li et al., 2012). |
| Tree-based models | Decision Tree [c] | A tree-like structure to make decisions based on a series of criterions. | Creates a hierarchical partitioning of the feature space based on information-theoretic criteria (e.g., entropy), offering transparent decision rules that facilitate model interpretation (Sundhari, 2011). |
| Ensemble models | Random Forest [c] | A group of decision trees where results are aggregated into one. | A group of decision trees act as weaker estimators, which can also output feature importance and therefore offer interpretability (Omondiagbe et al., 2019) and reduce variance. |
| | Adaptive Boosting (AdaBoost) [c] | Sequentially-built weaker estimators by strategically reweighting training data. | Sequentially builds an ensemble by focussing on previously-misclassified instances, thereby creating a complementary set of weaker learners that collectively achieve higher accuracy through targeted error correction (Wang et al., 2015). |
| | Bagging Estimator [c] | Independently-built weaker estimators where results are aggregated to predict continuous value. | Trains models on random subsets of data with replacement, therefore more robust predictions can be achieved (Chen and Ren, 2009). Inclusion stems from whether averaging predictions from multiple models with diverse training data improves results. |
| | Extreme Gradient Boosting [c] | Gradient boosting with regularisation and built-in cross validation. | Optimised Gradient Boosting algorithm which is known for scalability and performance through systematic tree pruning and parallel processing (Asselman et al., 2023). |
| Nearest neighbours | K-Nearest Neighbours [c] | K-Nearest Neighbours to predict continuous or binary value. | Predicts outcomes based on local patterns in the feature space by averaging values from distance-based training, providing a non-parametric alternative that captures complex decision boundaries (Tang et al., 2020). |
| Deep learning | Neural Network [c] | A general-purpose feed-forward neural network with hidden layers. | Fully-connected, feed-forward neural networks may enable us to capture complex nonlinear relationships, beyond the capacity of shallow learning algorithms (Wang, Wang, et al., 2020). |

[a] Regression-only models.
[b] Classification-only models.
[c] Models applicable for both regression and classification.

Table 6. Benchmarked Neural Network architectures for both regression and classification tasks

| ID | Study Origin | Architecture |
|---|---|---|
| 1 | Mondal et al. (2021) | Hidden layers: 2<br>Hidden units per layer: 8<br>Activation function: Sigmoid<br>Optimiser: Adaptive moment estimation (Adam) |
| 2 | Shen and Lee (2016) | Hidden layers: 3<br>Hidden units per layer: 512<br>Activation function: Rectified Linear Unit (ReLU)<br>Optimiser: Root mean square propagation (RMSProp) |
| 3 | Salminen et al. (2020) | Hidden layers: 2<br>Hidden units per layer: 128; 64<br>Activation function: ReLU<br>Optimiser: Adam |
| 4 | Custom implementation | Hidden layers: 3 (Shen and Lee, 2016)<br>Hidden units per layer: 64 (Salminen et al., 2020)<br>Activation function: Leaky ReLU (with 0.01 slope) (Braiek and Khomh, 2023)<br>Optimiser: SGD with Momentum (Omondiagbe et al., 2019) |

### *3.4.2 Feature Selection and Engineering*

In adherence to best practices (see Table 4), we commenced feature selection by manually inspecting redundancies before assessing the presence of multicollinearity. First, the metric *User Development Index* was excluded as a feature, as this metric is simply the sum of users' *Questions* and *Answers* – as established in prior work (Zolduoarrati et al., 2025c). As *Answers* constitutes the target variable for RQ1, incorporating this metric would introduce bias. This decision is further supported by our initial data exploration, which revealed a high Pearson's correlation coefficient ($r$) of 0.994 ($p < 0.05$) between *Answers* and *User Development Index*. This signifies that *User Development Index* is highly explanatory of the target variable, rendering its inclusion unnecessary.

Several diagnostic techniques are available to assess multicollinearity amongst features. A common initial step involves inspection of the correlation matrix to identify high pairwise correlations between features. However, the problem with this method is that the correlations do not necessarily mean multicollinearity, and that the arbitrary removal of a feature from each highly-correlated pair presents a discretionary approach that may introduce researcher bias. Therefore, more robust measures are necessary for definitive and scientific assessment. In this regard, our study employed Variance Inflation Factor (VIF) to assess multicollinearities. While there is no universally accepted threshold for VIF, values exceeding 5 are widely accepted (Shrestha, 2020), and thus features exhibiting a VIF score exceeding 5 were removed. This ensures that the features we incorporate into the subsequent models are statistically independent (i.e., do not overlap with others) and contribute uniquely to the prediction process. All available variables were subject to this assessment except the ones defined as target variables: users' *Answers* (RQ1), users' *Code Quality* violation densities (RQ2), and users' *Dropout* status (RQ3). Including these target variables in the VIF assessment would be redundant, as they represent the very outcomes we are seeking to predict. On the other hand, their exclusion from VIF assessments ensures that we solely evaluate the features that contribute explanatory power to the models. VIF was computed only once for each RQ, prior to evaluating all combinations of FE and modelling techniques, as depicted in Figure 1. Following VIF calculation across all RQs, we found severe multicollinearity among *User Management Index*, *UpVotes*, and *DownVotes*, whereby all three variables exhibited VIF exceeding 100. Further exploration using Pearson's $r$ confirmed significant correlations between *User Management Index* against both *UpVotes* ($r = 0.820$; $p < 0.01$) and *DownVotes* ($r = 0.754$; $p < 0.01$). Given that *User Management Index* was simply the sum of *UpVotes* and *DownVotes* (Zolduoarrati et al., 2025c), we opted to exclude it as it is redundant, and retained *UpVotes* and *DownVotes* as features. We believe such an approach addresses multicollinearity while preserving the informative aspects of user content curation activities.

The impact of FE varies greatly across different prediction tasks and model types, making empirical comparison necessary to avoid potentially counterproductive transformations that might discard meaningful variance (Strelcenia and Prakoonwit, 2023). Regarding FE, we shortlisted four different techniques to use, seen in Table 7 below. Following the selection of appropriate FE techniques, we then evaluated the combined effect of such techniques with different models to address RQ1, RQ2, and RQ3. For RQ2, there were 20 target variables as we have 4 quality dimensions across 5 languages. This resulted in varying totals of unique evaluations for each RQ, as shown in Table 8.

Table 7. FE techniques used

| Technique Name | Description | Merit for Inclusion |
|---|---|---|
| Standardisation | Rescales features to have a mean of 0 and standard deviation of 1. | Chosen for its ability to centre features around zero while preserving relative differences (Raju et al., 2020). This is valuable when features have varying scales (e.g., sentiment scores and readability indices both having different orders of magnitude). |
| Normalisation | Rescales features to be between 0 and 1. | Chosen for its ability to constrain all features to a consistent range, making it especially suitable for handling the diverse user metrics that may contain outliers (Raju et al., 2020). This is valuable when datasets contain both active users with hundreds of answers, and passive users with little to no activity. |
| Log-transformation | Log-transforms features by compressing large values and spreading smaller ones. | Chosen to address the typically skewed distributions in Stack Overflow activities (Feng et al., 2014; Zolduoarrati et al., 2024), where most users exhibit low to moderate activity levels, whilst only several users demonstrate high activity levels. |
| Power transformation | Transforms features by raising them to a specific power to achieve a more normal distribution. | Chosen to address non-linearities that might be missed by otherwise linear models (Carroll and Ruppert, 1981). Similar strength as log-transformation, but power transformation also works well with variables of ratio scale. |
| No feature engineering | Bypasses any FE altogether. | Chosen to provide a critical control condition that allows for validation of whether these preprocessing steps actually improve predictive performance, since FE may not always be necessary (Cabello-Solorzano et al., 2023). |

Table 8. Total unique evaluations for RQ1 to RQ3

| RQ# | Models | FE techniques | Target variable(s) | Total unique evaluations |
|---|---|---|---|---|
| RQ1 | 18 | 5 | 1 | 90 |
| RQ2 | 18 | 5 | 20 | 1,800 |
| RQ3 | 12 | 5 | 1 | 60 |

#### *3.4.3 Hyperparameter Optimisation and Validation*

To achieve the most effective configuration for all benchmarked models, we used Bayesian optimisation (BO) to automatically tune hyperparameters instead of conventional heuristics such as grid-searching (Anghel et al., 2018). Tree-Structured Parzen Estimator was employed as the sampling method, enabling efficient exploration of the hyperparameter search space (Rong et al., 2021). All optimisation processes were automated using the Optuna[12] AutoML framework, which has been shown to yield better performance compared to other frameworks (Akiba et al., 2019).

The effectiveness of our Bayesian-tuned hyperparameter selection was finally assessed through genetic algorithm-based (GA) hyperparameter tuning (Shanthi and Chethan, 2022). However, this more computationally-intensive approach was applied judiciously, focussing solely on the top five performing models and their corresponding FE combinations identified during the initial BO stage. The aim was not to compose another set of hyperparameters, but instead to assert whether those identified by the BO significantly deviated from those obtained through GA (i.e., falling below a 5% margin). We automated GA using the Neural Network Intelligence (NNI) toolkit[13], an open-source platform specifically designed for efficient hyperparameter optimisation. Configurations for GA-based tuning are provided within our replication package (Zolduoarrati et al., 2025a)[14]. Finally, to ensure robust and fair evaluation while adhering to best practices (Omondiagbe et al., 2022), each model/FE combination was executed 100 times using the identified optimal hyperparameter sets. This enabled us to report both the mean and standard deviation for each combination, ensuring that the outcomes of such repeated runs were similar (Omondiagbe et al., 2022). We could not make any assumptions about the normality of the data, therefore we used Kruskal-Wallis tests given its non-parametric nature (Omondiagbe et al., 2022). Afterwards, we conducted post-hoc pairwise Conover-Iman tests with appropriate Bonferroni adjustments.

[12] https://optuna.org

[13] https://github.com/microsoft/nni

[14] Replication package » Preprocessing and Model Selection » GA Configurations.docx

## 3.5 Textual-Based Predictions

### *3.5.1 Applicable Transformer-based Models*

Our selection of an appropriate transformer-based model for RQ4 involved a careful review of several popular options documented in recent literature. Key factors considered during this selection process included the models' reported performance on similar classification tasks, its model and token sizes, and the computational resources required for fine-tuning. We initially explored generative models prevalent in the NLP domain, such as Gemma and GPT-4, but these models are mainly used for text generation and reasoning-related tasks, which does not fit our classification objective. Moreover, fine-tuning these generative models for our specific use case would require significant resources without any guarantee of success. For instance, tuning GPT models with OpenAI APIs may require up to 160 GB of RAM and $12,000 (Bowen et al., 2024), and even then may still yield negligible performance on our dataset. Thus, we concluded that the investment of resources in fine-tuning state-of-the-art LLMs would not yield a commensurate return, given the uncertainty of the outcome (see Section 6).

We thus looked at BERT, which has been demonstrably successful in various text classification tasks (Sun et al., 2019). The BERT family models were preferred in our study due to their ease of fine-tuning, abundant online resources, and established track record within similar research (He et al., 2024). We did not explore other model families as that would involve significant trial and error cycles, and the purpose of using transformer-based models in our study was simply to validate the numeric-based model highlighted in RQ3 rather than conducting a comparative analysis of transformer models. We acknowledge that there may be other models capable of surpassing BERT's performance, which is discussed in Section 6. We did review sub-variants of BERT model family, including DistilBERT (a more lightweight version) and RoBERTa (known for its more optimised training regime) (Adoma et al., 2020). However, our data incorporates both natural language (post content) and programming code (accompanying code snippets). After a brief literature review, we found that similar works tend to employ transformer models capable of processing both natural and programming languages at the same time (Feng et al., 2020). It remains theoretically possible that transformer models trained solely on natural language could suffice for our specific use case, yet this scenario is considered less likely (see Section 6).

To address the challenge of processing both natural and programming languages, we opted for CodeBERT, a BERT variant developed by Microsoft specifically designed to handle such tasks (Feng et al., 2020). CodeBERT's frequent application in code-related tasks within the literature supports our belief that it should be effective at handling the specific challenges posed by our dataset (Mashhadi and Hemmati, 2021; Zhou et al., 2021), such as the presence of both code snippets and paragraphs within a single post and therefore should be treated as two separate components. While we acknowledge the existence of Stack Overflow-specific models like SOBert (He et al., 2024) and BERTOverflow (Tabassum et al., 2020), these models have seen limited application within relevant literature, which raises generalisability threats. Moreover, these models were shown to yield worse performance than the base BERT variant. In one instance, BERTOverflow demonstrated a worse performance over base BERT (3.9% difference in $F_1$-score), which suggests that BERTOverflow may not be a suitable model for various Stack Overflow-related processes (Yang et al., 2022). On the other hand, CodeBERT was proven to be generalisable to other tasks (Zhou et al., 2021). Thus, we picked CodeBERT to carry out RQ4.

### *3.5.2 Preprocessing*

Our text data preprocessing underwent a two-pronged approach, catering to the distinct characteristics of textual content within posts and code snippets, which was done by separating user-generated paragraphs from embedded code elements (see Section 3.2). For posts, we first conducted Unicode normalisation, ensuring different character representations were standardised to consistent canonical forms (Ansary et al., 2024). For instance, the character é may be represented in in Unicode in either a precomposed form representing the entire character (é) or a decomposed form that combines the letter (e) with an acute accent character (´). We used the NFC (Normalisation Form Canonical Composition) form due to its key characteristic of being non-destructive whilst preserving information (Öhman et al., 2023). We then applied standard NLP preprocessing techniques such as converting to lower case, and removing extraneous whitespace, stop words and punctuation, except for punctuation within links. This exception is grounded in research highlighting the importance of links shared by active users (Ahasanuzzaman et al., 2018). Given our goal of predicting user churn, retaining such textual features becomes crucial as they are valuable indicators of users remaining as active contributors. Figure 2 depicts an example of text before and after punctuation removal, highlighting that links are intentionally preserved.

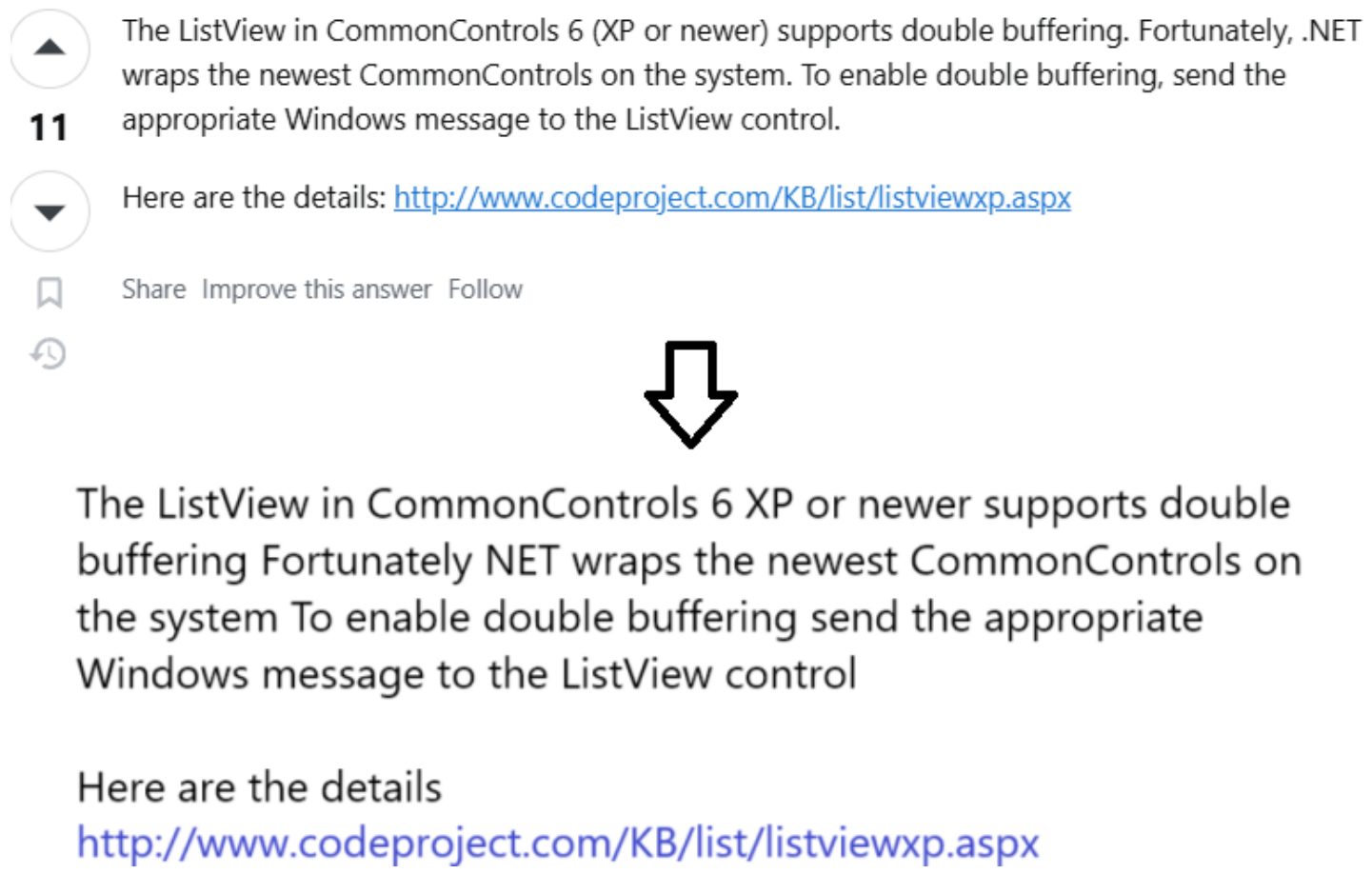

Figure 2. Example punctuation removal while preserving links

Code snippet preprocessing commenced with language identification (Feng et al., 2020). We also reflected upon the work by Zhang et al. (2022), who employed CodeBERT to generate Stack Overflow question titles. Initially, we explored leveraging question tags to classify the programming languages inside the snippets. Tags often indicate the programming language (e.g., *"Java"*) and the relevant technology (e.g., *"Maven"*). However, this approach was complicated by threads that encompassed multiple programming languages. For instance, Figure 3 depicts a question-answer pair regarding how to render HTML elements within a TextView component in the Android framework. While the primary question was explicitly tagged with HTML, the provided answers were in Java and Kotlin. To address this challenge, we used Guesslang[15], a tool that has previously yielded positive results in identifying code snippets on Stack Overflow (Zolduoarrati et al., 2025c). Guesslang is a deep learning tool that infers programming languages from plain text, and is a key component of popular developer tools such as Visual Studio Code[16]. We first performed a pilot reliability test for our extracted code blocks. 385 different snippets were randomly selected from the larger pool of 9,467,264 snippets, following Cochran's formula for finite population with a 5% margin of error and a 95% confidence level (Naing et al., 2006). Our pilot test yielded an accuracy of 84.67% (see our replication package (Zolduoarrati et al., 2025a)[17] for details), which highlights its reliability given that an accuracy exceeding 80% is generally considered acceptable (Chan and Paelinckx, 2008). Guesslang predictions were incorporated into a database column named *"GuesslangTag"*, and we then filtered the rows to retain only those where *"GuesslangTag"* matched the five studied languages: SQL, JavaScript, Python, Ruby, and Java. Table 9 documents the number of snippets associated with each Guesslang prediction. Moreover, it is worth noting that four of the five languages under study (JavaScript, Python, Ruby, and Java) were also previously utilised in the pre-training of CodeBERT (Feng et al., 2020), seen in Table 10. This prior exposure reinforces CodeBERT's applicability to our current study.

After extracting the relevant code snippets, we prioritised minimal intervention for subsequent preprocessing to limit the risk of introducing bias to subsequent analyses (Meldrum et al., 2020). This is all the more important as CodeBERT relies on analysing Abstract Syntax Trees (AST) generated from the input code (Feng et al., 2020). We thus argue that preserving the inherent structure of the code snippets is crucial to capture any latent information. To this end, our preprocessing techniques are restricted to the removal of inline comments and docstrings (e.g., Javadoc comments) as these elements can introduce noise into the data (Sharma et al., 2022). Our extraction pipeline uses a custom Python script with regular expressions to remove single-line and multi-line comments across all languages, as illustrated in Figure 4. Our replication package (Zolduoarrati et al., 2025a) includes demonstrations of the regular expressions[18] as well as the comment identifiers for each language[19].

[15] https://github.com/yoeo/guesslang
[16] https://code.visualstudio.com/updates/v1_60#_automatic-language-detection
[17] Replication package » Evaluation » Guesslang Samples.xlsx

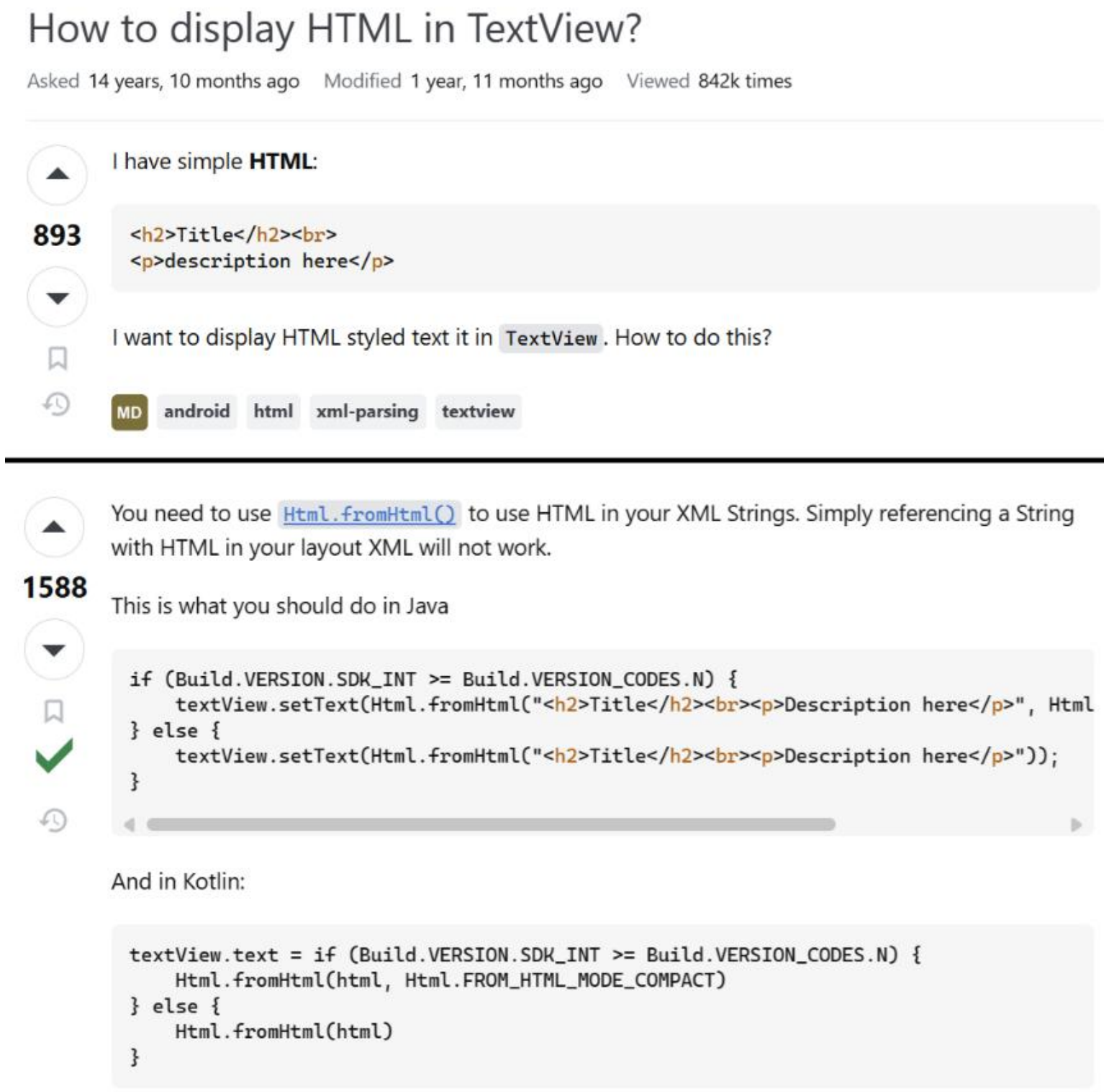

Figure 3. Example question-answer pair with differing languages

Table 9. Snippets per Guesslang prediction

| **Guesslang Predictions** | **Total Snippets** |
|---|---|
| SQL | 1,017,698 |
| JavaScript | 742,296 |
| Python | 506,532 |
| Ruby | 337,517 |
| Java | 322,193 |

Table 10. Intersection of studied languages and languages used to pre-train CodeBERT

| **Language** | **Studied in current and prior work (Zolduoarrati et al., 2025b)** | **Used to pre-train CodeBERT** |
|---|---|---|
| SQL | ✓ | ✗ |
| JavaScript | ✓ | ✓ |
| Python | ✓ | ✓ |
| Ruby | ✓ | ✓ |
| Java | ✓ | ✓ |
| PHP | ✗ | ✓ |
| Go | ✗ | ✓ |

[18] Replication package » Preprocessing and Model Selection » Regex Demonstrations.txt

[19] Replication package » Preprocessing and Model Selection » Comment Identifiers.pdf

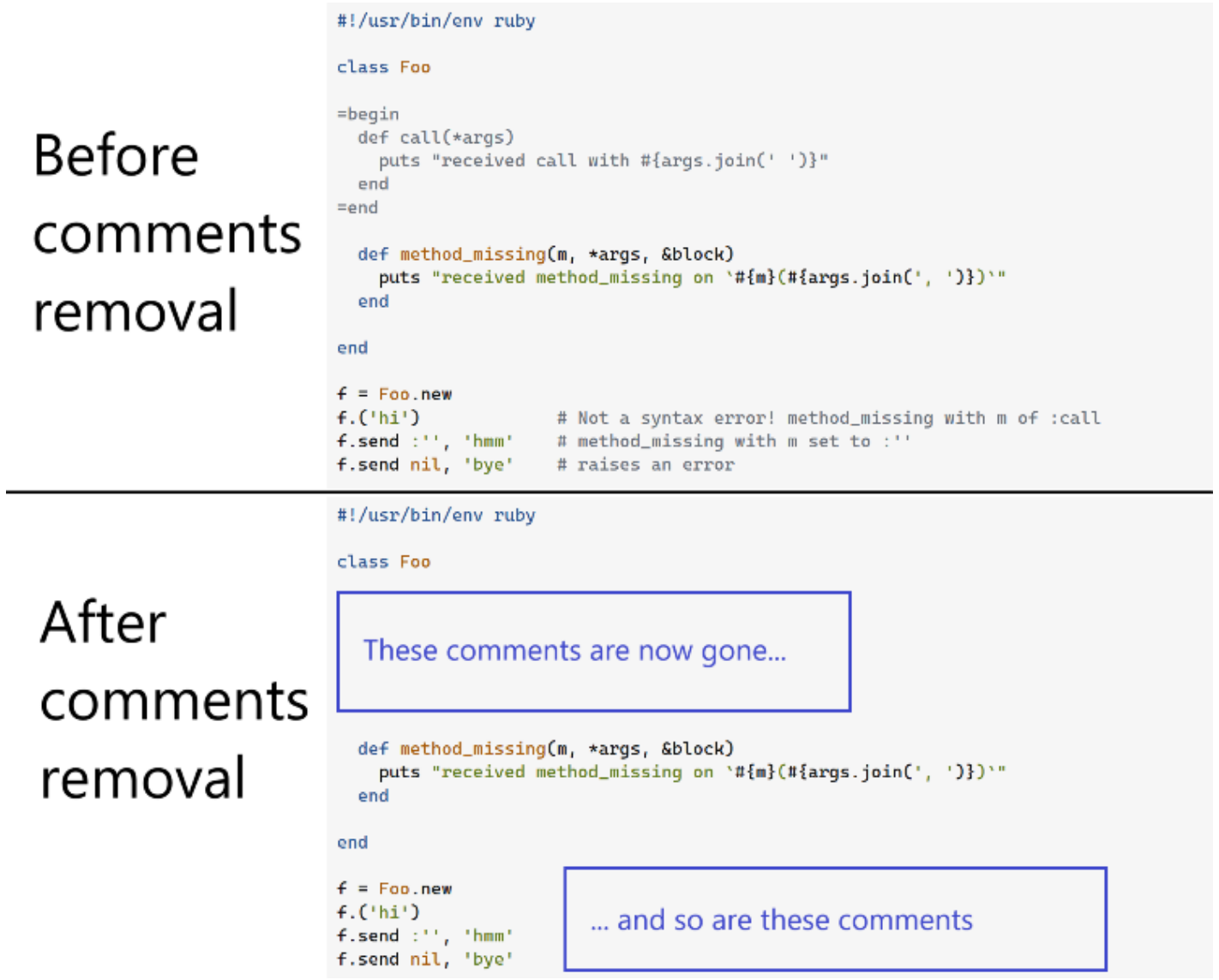

Figure 4. Example single and multi-line comments removal in Ruby

Finally, textual elements were segmented into individual units suitable for machine learning models. We added the special identifier tokens *[CLS]* and *[EOS]* to explicitly indicate the beginning and end of each sequence (Feng et al., 2020), which enabled the model to learn meaningful embeddings (Nguyen et al., 2023). To differentiate the two language segments (natural vs. programming language), a separator token *[SEP]* was inserted in the middle. Each language segment was then tokenised into individual word tokens. These tokens are denoted as $w_1, w_2, \ldots, w_n$ for the natural language portion and $c_1, c_2, \ldots, c_m$ for the programming language code (Feng et al., 2020).

We used the CodeBERT tokeniser to perform this step on both the post content and accompanying code snippets, which was streamlined using HuggingFace's *transformers*[20] library. The CodeBERT tokeniser produces two outputs: input identifiers and an attention mask, which are generated for both the paragraphs and the code snippets within each post. The original CodeBERT implementation restricts the maximum sequence length to 512 tokens. However, due to internal model requirements, only sequences up to 509 tokens can be processed directly given the existence of the aforementioned special tokens *[CLS]*, *[SEP]*, and *[EOS]* (Feng et al., 2020). Input sequences larger than this were thus truncated to make room for these special tokens. We acknowledge this as a limitation as this truncation could introduce minor inaccuracies, but such mechanisms should not significantly compromise the outcomes of fine-tuning processes (Mashhadi and Hemmati, 2021). We discuss this issue further in Section 6.

### 3.5.3 *Modelling*

We harnessed a pre-trained CodeBERT model readily available from the HuggingFace hub[21]. We used the February 12, 2022 version, which is the most recent repository commit at the time of writing. However, subsequent updates to CodeBERT may render this specific version suboptimal, which we discuss further in Section 6. Our study opted to use the default model configuration, aligning with the work by Mashhadi and Hemmati (Mashhadi and Hemmati, 2021) which showcased that the default model configuration can also yield good insights. Additionally, for transformer-based models, results obtained from optimal hyperparameter configurations are often closely aligned with those obtained from default settings (Fazeli, 2024). The core architecture consists of a Transformer decoder with six layers, each featuring a hidden state dimension of 768 and utilising 12 attention heads. For fine-tuning, the AdamW optimiser was used given its superior performance compared to other optimisers, such as standard Adam (Llugsi et al., 2021). We limited the training epoch to 8 and learning rate to be 1e-5, which have been shown to be sufficient to obtain convergence (Zhou et al., 2021) while aligning with the default settings employed by the CodeBERT authors (Feng et al., 2020). However, due to computational resource constraints, the batch size was set to 8 (Zhou et al., 2021). While larger batch sizes (e.g., 12 or 16) might potentially improve training efficiency, they resulted in local machine crashes during our experiments. The extracted contextual embeddings were subsequently processed by a fully connected layer for user *Dropout* classification, which performs a transformation of the embedding vectors, effectively compressing the data into a lower-dimensional representation suitable for the classification task (Zhou et al., 2021). Finally, a sigmoid activation function was applied to the output layer, generating values between 0 and 1. Values closer to 1 signify a higher likelihood of the input representing that the corresponding user would not drop out. Figure 5 depicts the CodeBERT input/output representation.

[20] https://github.com/huggingface/transformers

[21] https://huggingface.co/microsoft/codebert-base

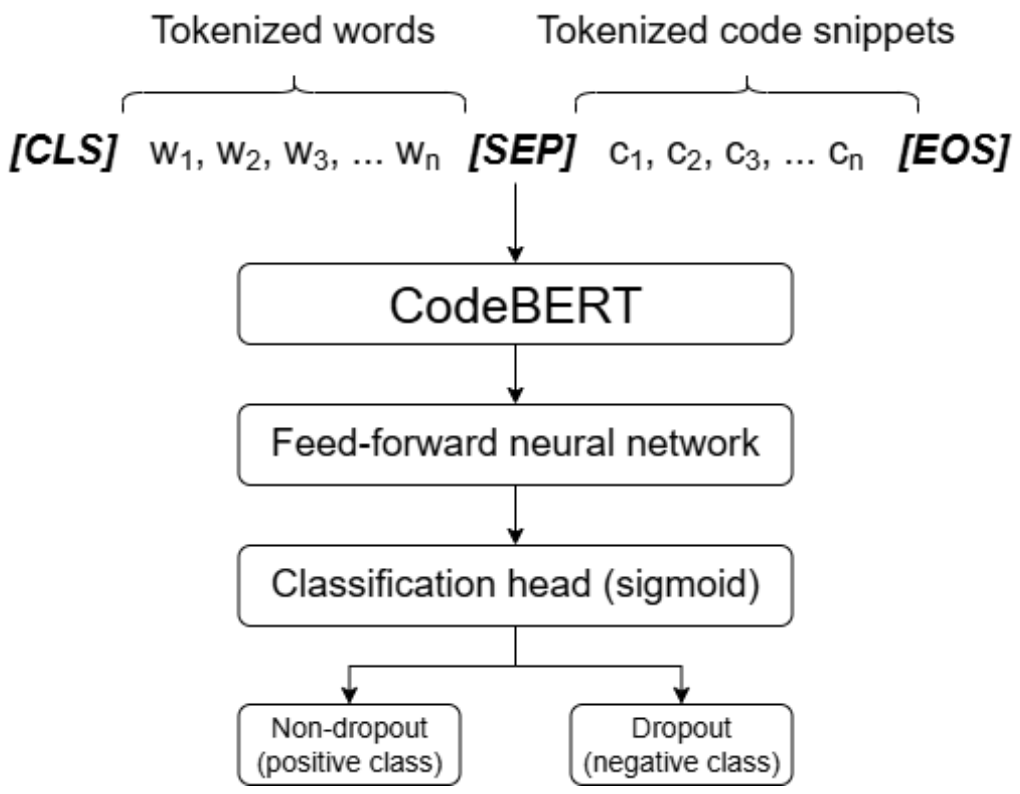

Figure 5. CodeBERT input/output representation

### 3.6 Evaluation

To comprehensively evaluate model performance, we adopted a selection of evaluation metrics tailored to the task at hand. For regression tasks (RQ1 and RQ2), we report the Root Mean Squared Error (RMSE) and $R^2$. For classification tasks (RQ3 and RQ4), we evaluated model performance using both accuracy and $F_1$-score. Classification using deep learning methods (i.e., Neural Networks in RQ3 and transformer-based models in RQ4) used the binary cross-entropy loss function.

To assess the efficacy of the numeric-based (RQ3) and textual-based (RQ4) *Dropout* prediction approaches, we benchmarked the outcomes of both models against a ‘ground truth’ dataset that sampled 385 users, their posts, and their *Dropout* status (giving a 95% confidence interval and 5% error margin). The *Dropout* statuses of users were derived from prior work (Zolduoarrati et al., 2025c). Accuracy and $F_1$-scores were then computed to see whether the prediction two approaches could offer complementary insights while mitigating each other’s limitations. For instance, transformer-based models might reveal dropout tendencies that are not readily apparent in purely numerical measurements. The results of this ground truth analysis may be seen in our replication package (Zolduoarrati et al., 2025a)[22], and are also discussed briefly in Sections 4.3. and 4.4. Due to file size limitations, we were unable to include the final models for each RQ within the replication package. Instead, the shallow learning models (RQ1 to RQ3) have been uploaded online for accessibility[23], following literature best practices (Omondiagbe et al., 2022). The fine-tuned model (RQ4) has been stored in a public Hugging Face repository[24].

## 4 RESULTS

The computational environment for this study leveraged a local system equipped with a Ryzen 9 5950x CPU, Nvidia RTX 3090 GPU, and 128 GB of DDR4 memory at 3.6 GHz. We overclocked the GPU from 1,395 to 1,695 MHz, and the CPU from 3.4 to 4.7 GHz, both figures being the most stable after numerous trials. All 32 available CPU threads were utilised to achieve concurrent execution whenever possible. Prior to answering the RQs, we present a high-level overview of the strengths and weaknesses of the various numeric-based models used to answer RQ1, RQ2, and RQ3 (see Table 5). This evaluation considers factors such as wall time required to both train with default hyperparameters and to conduct BO, as well as the overall performance across classification and regression tasks. Practitioners may reflect on each model’s computational efficiency, required time investment, and performance with respect to their specific needs. Table 11 presents the summarised average performance metrics across 100 runs for all RQs and FE techniques employed. For instance, Random Forest with an $R^2$ value of 0.44 ± 0.159 signifies that the Random Forest model achieved an average $R^2$ of 0.44 with a standard deviation of 0.159, across both RQ1 and RQ2 (regression tasks), and across all FE methods applied. Our replication package includes the total wall time to train each model combination without tuning (Zolduoarrati et al., 2025a)[25].

[22] Replication package » Results » Binary Classification Benchmarks.xlsx

[23] https://tinyurl.com/47yc9xju

[24] https://huggingface.co/modelling101/CodeBERT-SO

### 4.1 Users' Answers (RQ1)

This section presents results for RQ1: *What is the best modelling approach to predict users' answers?*

To address multicollinearity among the predictor variables, we calculated the VIF for each feature as described in Section 3.4. Appendix A presents the VIF values for each feature, and those with VIFs exceeding 5 were excluded from the final model to mitigate the influence of collinearity. This resulted in a reduced set of 15 features from the initial pool of 20.

Notably, the 21 algorithms presented in both Table 5 and Table 11 encompass regression-only models (e.g., OLS), classification-only models (e.g., Logistic Regression), and hybrid models capable of both regression and classification tasks (e.g., Random Forests). However, for RQ1 where the focus is to compare models for predicting users' *Answers* (i.e., total questions that a user is likely to answer), the evaluation will be restricted to a subset of 17 algorithms, which are either regression-only models or hybrid ones. Consequently, results herein are presented for the aforementioned 17 models rather than the full set of 21, shown with respect to each model/FE combination in Table 12. Values following the ± sign indicate standard deviation, whereas those enclosed in brackets indicate the median across all 100 repeated runs. The shallow baseline model (first row of Table 12) exhibited the same performance across all FE techniques, with an $R^2$ of -0.00003 and an RMSE of 173.32 for all 100 runs. In contrast, the deep Neural Network baseline model demonstrated sensitivity to the preprocessing methods applied, shown in the second-to-last row of Table 12.

We observed that the combination of standardisation for FE and the Bagging (bootstrap aggregating) estimator achieved the best performance across all tested combinations (highlighted in **bold** in Table 12), with $R^2 = 0.821 \pm 0.098$ indicating that the model can explain 81-83% of the variance in the number of users' *Answers*. Moreover, the RMSE of 0.331 ± 0.053 suggests an average prediction error of 0.331 units on the number of answers for each user. In other words, for a user with 10 *Answers*, the model's prediction may fall between 9.616 and 10.384. We argue such predictions are a good fit as the error margin falls relatively close to the true values (3.84%) and within 5% (Coulston et al., 2016). Our BO identified the optimal hyperparameters needed to obtain the aforementioned results, as depicted in Table 13. As previously outlined in Section 3.4.3, these hyperparameters were then validated using GA to ensure that the hyperparameter values obtained from both methods did not deviate by more than 5%, thereby ensuring that the values derived are correct.

Kruskal-Wallis tests were conducted to measure differences in performance ($R^2$ and RMSE) across the various model/FE combinations. We found statistically significant differences between all model/FE combinations (H = 7,111.137, $p < 0.01$ for $R^2$; H = 7,199.81, $p < 0.01$ for RMSE). Post-hoc Conover-Iman tests with Bonferroni corrections on pairwise differences between the model and FE combinations also indicated statistically significant differences ($p < 0.01$) in both $R^2$ and RMSE for all comparisons.

We found limitations with the Epsilon SVM model when applied to raw, unprocessed features. This combination resulted in errors where the model was unable to determine finite dual coefficients or intercepts, indicating that the input data contained wide-ranging values that exceeded the model's handling capabilities. This is indicated by 'N/A' in Table 12.

Our results also revealed that complex Neural Network models did not necessarily always outperform simpler shallow learning methods in predicting users' *Answers*. In fact, several shallow learning approaches, such as Extreme Gradient Boosting ($R^2 = 0.716$) and SGD ($R^2 = 0.736$), achieved superior performance compared to the Neural Network ($R^2 = 0.642$) when applied to features without prior scaling. This disparity aligns with established research indicating challenges faced by Neural Networks when handling large integers (Kloberdanz et al., 2022). Our results also highlight a contrast between Extreme Gradient Boosting and Bagging ensembles in terms of their performance consistency across different FE techniques. Specifically, while Bagging boasts a higher peak performance ($R^2 = 0.821$), its performance varies across FE techniques, going as low as $R^2 = 0.062$ for log-transformed features. In contrast, Extreme Gradient Boosting has more consistent performance across the board, achieving $R^2$ values between 0.694 and 0.716. These findings are corroborated by the standard deviations of the $R^2$ values for each model/FE combination. The standard deviations for Extreme Gradient

[25] Replication package » Results » Runtime for Models.xlsx

Boosting are consistently zero across all FE methods, whereas the standard deviations for Bagging range from 0.021 to 0.098, indicating greater variability. Our experiments thus suggest Extreme Gradient Boosting may be the most suitable modelling approach for tasks requiring consistent performance, especially when feature rescaling or transformation methods are not applicable due to environmental constraints.

### 4.2 Users' Code Quality Violations (RQ2)

This section presents results for RQ2: *What is the best modelling approach to predict users' code quality?*

Similar to RQ1, we calculated the VIF for each feature to address multicollinearity among the predictor variables. The *Answers* variable, which serves as the target variable in RQ1, is now incorporated as a feature to predict the density of users' *Code Quality* violations, as previously outlined in Section 3.4. The VIF values for all features employed in RQ2 are documented in Appendix A, where those exceeding 5 were excluded. This resulted in a streamlined feature set containing 16 features, down from the initial 21.

We evaluated both deep and shallow baseline models against all 20 target variables, encompassing four quality dimensions across five programming languages. Due to the volume of results, Table 14 presents only the best-performing model/FE combination for each target variable based on its $R^2$, along with the corresponding hyperparameter sets used to achieve these results. The hyperparameter sets were all validated using GA, ensuring that the obtained hyperparameters do not deviate further than 5%. A comprehensive overview of all results can be found in the replication package (Zolduoarrati et al., 2025a)[26].

With regards to models, Table 14 reveals that the SGD regressor is the most frequent top performer, achieving the best results at predicting five *Code Quality*-related features: SQL Reliability, SQL Security, JavaScript Reliability, Python Reliability, and Ruby Performance. Bagging and Epsilon SVM came second equal, both achieving best performance in three *Code Quality* features. Bagging excelled at predicting SQL Performance, JavaScript Readability, and Java Performance violation density, while Epsilon SVM proved most effective for Python Readability, Python Security, and Ruby Security. Looking at individual languages, SGD performed well for SQL, particularly in the dimensions of Reliability ($R^2 = 0.674$) and Security ($R^2 = 0.682$). For Python, the Epsilon SVM model was the most performant, particularly in predicting Readability ($R^2 = 0.698$) and Security ($R^2 = 0.666$). However, there was no clear best-performing model for JavaScript, Ruby, or Java, as different models excelled in different quality dimensions. For example, each of the four JavaScript dimensions had a different model (SGD, Bagging, Bayesian, and AdaBoost) achieve the highest performance.

Table 11. Average training time, optimisation time, and performance for numeric-based models (100 runs)

| Umbrella Method | Method Name | Training Time (seconds) | | Optimisation | Performance | |
|---|---|---|---|---|---|---|
| | | Regression | Classification | Time | Regression ($R^2$) | Classification ($F_1$) |
| Linear models | OLS [a] | 0.260 | - | 6s | 0.029 ± 0.128 | - |
| | Logistic Regression [b] | - | 2.223 | 7m 59s | - | 0.662 ± 0.033 |
| | ElasticNet [a] | 0.119 | - | 9s | 0.482 ± 0.136 | - |
| | Ridge Classifier [b] | - | 0.324 | 54s | - | 0.652 ± 0.039 |
| | LassoLARS [a] | 0.080 | - | 11 s | -0.015 ± 0.187 | - |
| | SGD | 1.514 | 1.986 | 27m 32s | 0.462 ± 0.138 | 0.628 ± 0.046 |
| Bayesian models | Bayesian [a] | 0.337 | - | 29m | 0.469 ± 0.125 | - |
| | ARD [a] | 0.213 | - | 25m | 0.029 ± 0.129 | - |
| Outlier-robust models | Huber [a] | 12.832 | - | 16m 34s | 0.475 ± 0.113 | - |
| | Theil-Sen [a] | 126.733 | - | 7h 4m | 0.002 ± 0.103 | - |
| SVM | Epsilon SVM [a] | 530.943 | - | 11h 1m | 0.445 ± 0.159 | - |
| | Nu SVM [a] | 1,008.566 | - | 15h 24m | 0.027 ± 0.095 | - |
| | Linear SVM | 169.477 | 155.809 | 2h 41m | -0.500 ± 2.317 | 0.694 ± 0.019 |
| | C-SVM [b] | - | 1,118.077 | 12h 32m | - | 0.662 ± 0.013 |
| Tree-based models | Decision Tree | 3.424 | 5.787 | 1m 35s | 0.018 ± 0.066 | 0.733 ± 0.012 |
| Ensemble models | Random Forest | 238.921 | 95.439 | 4h 14m | 0.440 ± 0.139 | 0.734 ± 0.001 |
| | AdaBoost | 28.659 | 32.787 | 2h 1m | 0.403 ± 0.230 | 0.665 ± 0.013 |
| | Bagging | 24.458 | 44.252 | 4h 22m | 0.473 ± 0.130 | 0.659 ± 0.010 |
| | Extreme Gradient Boosting | 1.065 | 1.370 | 1h 27m | 0.443 ± 0.128 | 0.825 ± 0.000 |
| Nearest neighbours | K-Nearest Neighbours | 2.299 | 0.104 | 10h 7m | 0.487 ± 0.115 | 0.737 ± 0.007 |
| Deep learning | Neural Network | - | - | 21m | 0.033 ± 0.110 | 0.895 ± 0.019 |

[a] Regression-only models.
[b] Classification-only models.

[26] Replication package » Results » RQ2

Table 12. Benchmarking results for RQ1 (100 runs)

| **Model Name** | **Log-transformation** | | **No feature engineering** | | **Normalisation** | | **Power-transformation** | | **Standardisation** | |
|---|---|---|---|---|---|---|---|---|---|---|
| | **$R^2$** | **RMSE** | **$R^2$** | **RMSE** | **$R^2$** | **RMSE** | **$R^2$** | **RMSE** | **$R^2$** | **RMSE** |
| Shallow Baseline | -0.00003 ± 0.0 (-0.00003) | 173.32 ± 0.0 (173.32) | -0.00003 ± 0.0 (-0.00003) | 173.32 ± 0.0 (173.32) | -0.00003 ± 0.0 (-0.00003) | 173.32 ± 0.0 (173.32) | -0.00003 ± 0.0 (-0.00003) | 173.32 ± 0.0 (173.32) | -0.00003 ± 0.0 (-0.00003) | 173.32 ± 0.0 (173.32) |
| AdaBoost | -1.59 ± 4.399 (0.559) | 20.889 ± 0.721 (0.644) | 0.549 ± 0.027 (0.568) | 4.634 ± 0.02 (0.638) | 0.349 ± 0.009 (0.542) | 4.093 ± 0.007 (0.656) | 0.004 ± 0.037 (0.599) | 4.756 ± 0.027 (0.614) | 0.258 ± 10.999 (0.171) | 15.607 ± 1.423 (1.033) |
| ARD | 0.186 ± 0.0 (0.216) | 8.636 ± 0.0 (0.858) | 0.716 ± 0.0 (0.721) | 2.238 ± 0.0 (0.512) | 0.716 ± 0.0 (0.721) | 2.223 ± 0.0 (0.512) | 0.081 ± 0.0 (0.058) | 9.609 ± 0.0 (0.941) | 0.716 ± 0.0 (0.721) | 2.231 ± 0.0 (0.512) |
| Bagging | 0.062 ± 0.027 (0.72) | 5.721 ± 0.024 (0.513) | 0.816 ± 0.088 (0.73) | 0.437 ± 0.066 (0.51) | 0.229 ± 0.021 (0.732) | 9.638 ± 0.019 (0.506) | 0.815 ± 0.046 (0.699) | 0.452 ± 0.099 (0.54) | **0.821 ± 0.098 (0.709)** | **0.331 ± 0.053 (0.61)** |
| Bayesian | 0.185 ± 0.0 (0.216) | 8.637 ± 0.0 (0.859) | 0.716 ± 0.0 (0.721) | 2.238 ± 0.0 (0.512) | 0.716 ± 0.0 (0.721) | 2.238 ± 0.0 (0.513) | 0.081 ± 0.0 (0.058) | 9.609 ± 0.0 (0.941) | 0.716 ± 0.0 (0.721) | 2.239 ± 0.0 (0.512) |
| Decision Tree | 0.138 ± 0.002 (0.078) | 9.091 ± 0.001 (0.931) | 0.255 ± 0.005 (0.163) | 7.954 ± 0.003 (0.887) | 0.253 ± 0.004 (0.163) | 7.976 ± 0.002 (0.887) | 0.258 ± 0.047 (0.055) | 7.927 ± 0.025 (0.943) | 0.253 ± 0.002 (0.163) | 7.976 ± 0.001 (0.887) |
| ElasticNet | 0.186 ± 0.0 (0.218) | 8.635 ± 0.0 (0.858) | 0.716 ± 0.0 (0.721) | 2.238 ± 0.0 (0.512) | 0.698 ± 0.001 (0.6) | 2.527 ± 0.001 (0.613) | 0.081 ± 0.0 (0.058) | 9.609 ± 0.0 (0.941) | 0.733 ± 0.0 (0.729) | 1.95 ± 0.0 (0.505) |
| Epsilon SVM | 0.061 ± 0.0 (-12.872) | 9.79 ± 0.0 (3.612) | N/A | N/A | -0.255 ± 0.0 (-10.35) | 12.408 ± 0.0 (3.267) | -0.151 ± 0.0 (-6.545) | 11.588 ± 0.0 (2.664) | -0.096 ± 5.622 (-3.311) | 11.137 ± 1.226 (1.44) |
| Extreme Gradient Boosting | 0.712 ± 0.0 (0.556) | 2.294 ± 0.0 (0.646) | 0.716 ± 0.0 (0.72) | 2.237 ± 0.0 (0.513) | 0.714 ± 0.0 (0.585) | 2.265 ± 0.0 (0.625) | 0.694 ± 0.0 (0.613) | 2.585 ± 0.0 (0.603) | 0.696 ± 0.0 (0.685) | 2.545 ± 0.0 (0.544) |
| Huber | 0.176 ± 0.0 (0.191) | 8.73 ± 0.0 (0.872) | 0.735 ± 0.0 (0.722) | 1.915 ± 0.0 (0.512) | 0.727 ± 0.0 (0.713) | 2.05 ± 0.0 (0.52) | 0.078 ± 0.0 (0.055) | 9.634 ± 0.0 (0.943) | 0.726 ± 0.0 (0.727) | 2.065 ± 0.0 (0.507) |
| K-Nearest Neighbours | 0.779 ± 0.0 (0.673) | 1.14 ± 0.0 (0.555) | 0.817 ± 0.049 (0.763) | 0.41 ± 0.009 (0.478) | 0.491 ± 0.096 (0.66) | 5.358 ± 0.085 (0.594) | 0.606 ± 0.097 (0.683) | 3.874 ± 0.001 (0.602) | 0.787 ± 0.013 (0.658) | 0.99 ± 0.092 (0.487) |
| LassoLARS | 0.186 ± 0.0 (0.217) | 8.635 ± 0.0 (0.858) | 0.716 ± 0.0 (0.721) | 2.238 ± 0.0 (0.513) | 0.727 ± 0.0 (0.666) | 2.055 ± 0.0 (0.56) | 0.081 ± 0.0 (0.058) | 9.609 ± 0.0 (0.941) | 0.731 ± 0.0 (0.727) | 1.986 ± 0.0 (0.506) |
| Linear SVM | 0.08 ± 0.001 (0.11) | 9.623 ± 0.001 (0.915) | 0.378 ± 0.239 (0.548) | 6.667 ± 0.138 (0.652) | 0.046 ± 0.0 (0.119) | 9.921 ± 0.0 (0.91) | 0.03 ± 0.001 (0.027) | 10.064 ± 0.0 (0.956) | 0.724 ± 0.0 (0.73) | 2.1 ± 0.0 (0.504) |
| OLS | 0.185 ± 0.0 (0.216) | 8.638 ± 0.0 (0.859) | 0.716 ± 0.0 (0.721) | 2.239 ± 0.0 (0.512) | 0.716 ± 0.0 (0.721) | 2.238 ± 0.0 (0.513) | 0.081 ± 0.0 (0.058) | 9.609 ± 0.0 (0.941) | 0.695 ± 0.0 (0.712) | 2.572 ± 0.0 (0.52) |
| Random Forest | 0.073 ± 0.0 (0.041) | 9.682 ± 0.0 (0.95) | 0.074 ± 0.001 (0.07) | 9.678 ± 0.045 (0.859) | 0.073 ± 0.0 (0.041) | 9.681 ± 0.0 (0.95) | 0.072 ± 0.0 (0.041) | 9.688 ± 0.0 (0.95) | 0.072 ± 0.094 (-0.063) | 9.692 ± 0.05 (0.905) |
| SGD | 0.185 ± 0.0 (0.216) | 8.639 ± 0.0 (0.858) | 0.736 ± 0.003 (0.703) | 1.903 ± 0.003 (0.529) | 0.716 ± 0.001 (0.72) | 2.234 ± 0.001 (0.513) | 0.081 ± 0.0 (0.058) | 9.609 ± 0.0 (0.941) | 0.684 ± 0.432 (0.688) | 2.741 ± 0.187 (0.542) |
| Theil-Sen | 0.187 ± 0.0 (0.22) | 8.627 ± 0.0 (0.857) | 0.43 ± 0.05 (0.247) | 10.766 ± 0.073 (0.959) | 0.746 ± 0.111 (0.278) | 1.736 ± 0.025 (0.822) | 0.081 ± 0.001 (0.314) | 9.606 ± 0.034 (0.88) | 0.042 ± 0.075 (0.175) | 1.532 ± 0.036 (0.896) |
| Neural Network Baseline | -0.021 ± 0.279 (-0.022) | 78.323 ± 0.008 (78.801) | -0.309 ± 0.207 (-0.316) | 5.796 ± 0.074 (5.517) | -0.019 ± 0.299 (-0.018) | 85.218 ± 0.258 (87.78) | -0.019 ± 0.052 (-0.018) | 93.035 ± 0.162 (90.551) | -0.02 ± 0.041 (-0.019) | 44.703 ± 0.166 (46.076) |
| Neural Network | 0.348 ± 0.266 (0.363) | 175.11 ± 0.253 (173.146) | 0.642 ± 0.107 (0.655) | 198.29 ± 0.037 (205.499) | 0.228 ± 0.273 (0.218) | 174.93 ± 0.084 (179.253) | 0.08 ± 0.187 (0.082) | 174.99 ± 0.176 (175.974) | 0.788 ± 0.005 (0.808) | 175.04 ± 0.291 (168.677) |

Values are presented as mean ± standard deviation (median)

Table 13. Optimal hyperparameter sets for standardisation and Bagging estimator

| Hyperparameter | Value from BO | Value from GA | Difference (%) |
|---|---|---|---|
| Total estimators | 532 | 543 | 2.026 |
| Max. sampling rate | 0.093 | 0.089 | 4.037 |
| Max. feature inclusion ratio | 0.821 | 0.807 | 1.649 |
| Bootstrap aggregation | True | True | - |
| Feature bootstrapping | False | False | - |

As with RQ1, we undertook Kruskal-Wallis tests to investigate differences in performance across various model/FE combinations. Tests were conducted once per each target variable, resulting in 20 unique tests with respect to $R^2$ and RMSE. We found statistically significant differences ($p < 0.01$) between all model/FE combinations regardless of which *Code Quality* aspect is being predicted. The H-statistic ranged from 5,211.21 to 7,114.95, further supporting rejection of the null hypothesis of equal performance. Post-hoc Conover-Iman tests with Bonferroni correction also showed statistically significant differences ($p < 0.01$) in both $R^2$ and RMSE values for all pairwise comparisons. There was no significant performance overlap between the different model/FE combinations, suggesting that each combination has its own merit and that no single combination consistently outperformed all others. However, certain combinations were found to be the most performant in predicting at least two target variables. For instance, log-transformation paired with AdaBoost exhibited the highest predictive performance for JavaScript Security and Java Reliability. Another example is power-transformation paired with Huber, being the most performant to model Python Performance and Ruby Reliability. Other models are less generalisable, being the most performant to model only a single *Code Quality* variable, seen in Table 14. For example, ElasticNet excelled at predicting Java Security violations without prior feature scaling, as well as Bayesian Ridge and K-Nearest Neighbours at predicting JavaScript Performance and SQL Readability when both algorithms were paired with normalisation.

Ensemble models were observed to perform well at predicting readability violations. Table 14 reveals that Extreme Gradient Boosting (coupled with power-transformation) and Random Forest (coupled with standardisation) emerged as the best performing models for predicting Ruby and Java readability violation densities, respectively. Interestingly, Neural Networks do not appear at all in Table 14. This suggests that even the greater complexity of Neural Networks may not be sufficient enough to capture more latent factors that drive or deter *Code Quality* violations. Alternatively, it could indicate that the current feature set may not be comprehensive enough, or that the key factors influencing these violations are already adequately captured by simpler models. However, an important caveat to consider is that all shallow learning models underwent a rigorous hyperparameter tuning process using BO, followed by further validation with GA. Neural Networks did not benefit from the same level of tuning, which may have contributed to the observed performance gap. This limitation is discussed in more detail in Section 6.

.

Encouragingly, all 20 prediction tasks shown in Table 14 achieved $R^2$ values exceeding 0.65, signifying that the employed models may explain at least 65% of the variability of each target variable. In the domain of social sciences, an $R^2$ of 0.5 is generally considered indicative of a good model fit (Ozili, 2023), but the computing domain typically sets a higher threshold of around 0.7 (Santos et al., 2022), which our $R^2$ values failed to reach (see Section 6). Despite this shortcoming, we still consider the obtained $R^2$ values to be relatively good, especially considering that predicting human behaviour is inherently complex (Schroeder, 2009), coupled with the limited number of predictor variables.

### 4.3 Users' Numeric-Based Dropout (RQ3)

This section presents results for RQ3: *What is the best modelling approach to predict users' dropout?*

As previously outlined in Section 3.4, all 20 *Code Quality*-related variables (i.e., target variables in RQ2) are incorporated as an expanded feature set for the exploration of RQ3, producing a final set of 41 features. Consistent with the approaches adopted in the first two RQs, we calculated the VIF for each feature in RQ3 to detect multicollinearity among the predictor variables (see Appendix A). We excluded features with VIFs over 5, thus reducing the number of features to 34. Consistent to the results shown in Sections 4.1 and 4.2, this RQ focusses on classifying user *Dropout*. Thus, a subset of 12 models was selected from the initial pool of 21, which includes classification-only models and hybrid models capable of performing both regression and classification. Table 15 summarises the accuracy and $F_1$-score metrics for each model/FE combination.

Table 14. Best FE/model combination for each RQ2 target variable (language + dimension) across 100 runs.

| Language | Dimension (Violation Density) | Feature Engineering | Model | $R^2$ | RMSE | Optimal Hyperparameter Set |
|---|---|---|---|---|---|---|
| SQL | Reliability | No feature engineering | SGD | 0.674 ± 0.253 (0.663) | 0.41 ± 0.017 (0.414) | Regularisation strength: 8.44<br>L1 penalty ratio: 0.24<br>Maximum no. of epochs: 123<br>Initial learning rate: 0.382 |
| | Readability | Normalisation | K-Nearest Neighbours | 0.691 ± 0.045 (0.678) | 0.425 ± 0.134 (0.43) | No. of neighbours: 100<br>Weight function: Uniform<br>Neighbour traversal: BallTree<br>Maximum leaf size: 23<br>Distance metric: Minkowski |
| | Performance | Power-transformation | Bagging | 0.67 ± 0.158 (0.69) | 0.556 ± 0.04 (0.532) | No. of weaker estimators: 569<br>Max. sampling proportion: 0.043<br>Max. features used: 0.235<br>Out-of-bag computation: False |
| | Security | No feature engineering | SGD | 0.682 ± 0.131 (0.65) | 0.601 ± 0.05 (0.614) | Regularisation strength: 6.835<br>L1 penalty ratio: 0.745<br>Maximum no. of epochs: 188<br>Initial learning rate: 0.781 |
| JavaScript | Reliability | Power-transformation | SGD | 0.657 ± 0.116 (0.645) | 0.409 ± 0.166 (0.399) | Regularisation strength: 59.471<br>L1 penalty ratio: 0.494<br>Maximum no. of epochs: 652<br>Initial learning rate: 3.439 |
| | Readability | Log-transformation | Bagging | 0.699 ± 0.198 (0.707) | 0.395 ± 0.255 (0.395) | No. of weaker estimators: 417<br>Max. sampling proportion: 0.142<br>Max. features used: 0.373<br>Out-of-bag computation: True |
| | Performance | Normalisation | Bayesian | 0.683 ± 0.072 (0.649) | 0.401 ± 0.243 (0.406) | Maximum no. of epochs: 861<br>Noise variance prior (shape): 3.604<br>Noise variance prior (scale): 0.046<br>Regularisation prior (shape): 0.358<br>Regularisation prior (scale): 9.986<br>Log marginal likelihood computation: False |
| | Security | Log-transformation | AdaBoost | 0.688 ± 0.045 (0.704) | 0.383 ± 0.027 (0.392) | No. of weaker estimators: 275<br>Learning rate: 0.324 |
| Python | Reliability | Standardisation | SGD | 0.693 ± 0.235 (0.688) | 0.539 ± 0.27 (0.548) | Regularisation strength: 1.418<br>L1 penalty ratio: 0.937<br>Maximum no. of epochs: 334<br>Initial learning rate: 2.163 |
| | Readability | Log-transformation | Epsilon SVM | 0.698 ± 0.074 (0.684) | 0.33 ± 0.15 (0.332) | Kernel function: radial basis function<br>Regularisation strength: 1.576<br>Epsilon insensitivity: 0.584<br>Maximum no. of epochs: 878 |
| | Performance | Power-transformation | Huber | 0.686 ± 0.219 (0.671) | 0.473 ± 0.096 (0.46) | Outlier sensitivity: 11.798<br>Maximum no. of epochs: 118<br>L2 regularisation strength: 6.337 |
| | Security | Power-transformation | Epsilon SVM | 0.666 ± 0.268 (0.691) | 0.651 ± 0.25 (0.626) | Kernel function: radial basis function<br>Regularisation strength: 1.124<br>Epsilon insensitivity: 0.99<br>Maximum no. of epochs: 785 |
| Ruby | Reliability | Power-transformation | Huber | 0.691 ± 0.075 (0.659) | 0.397 ± 0.094 (0.4) | Outlier sensitivity: 14.971<br>Maximum no. of epochs: 746 |

| Language | Dimension (Violation Density) | Feature Engineering | Model | $R^2$ | RMSE | Optimal Hyperparameter Set |
|---|---|---|---|---|---|---|
| | | | | | | L2 regularisation strength: 1.053 |
| | Readability | Power-transformation | Extreme Gradient Boosting | 0.688 ± 0.263 (0.714) | 0.413 ± 0.251 (0.43) | Base learner: gradient boosting trees<br>Max. tree depth: 3<br>Max. sampling proportion: 0.47<br>Learning rate: 0.054<br>No. of weaker estimators: 663<br>L1 regularisation strength: 0.516<br>L2 regularisation strength: 89 |
| | Performance | Log-transformation | SGD | 0.696 ± 0.168 (0.721) | 0.506 ± 0.06 (0.53) | Regularisation strength: 41.861<br>L1 penalty ratio: 0.654<br>Maximum no. of epochs: 541<br>Initial learning rate: 2.124 |
| | Security | Normalisation | Epsilon SVM | 0.69 ± 0.021 (0.693) | 0.428 ± 0.295 (0.436) | Kernel function: polynomial<br>Degrees: 2<br>Independent term: 7.972<br>Regularisation strength: 3.269<br>Epsilon insensitivity: 0.088<br>Maximum no. of epochs: 918 |
| Java | Reliability | Log-transformation | AdaBoost | 0.698 ± 0.143 (0.675) | 0.314 ± 0.033 (0.326) | No. of weaker estimators: 312<br>Learning rate: 0.234 |
| | Readability | Standardisation | Random Forest | 0.688 ± 0.272 (0.686) | 0.534 ± 0.119 (0.538) | No. of weaker estimators: 749<br>Max. tree depth: 3<br>Min. samples split: 0.357<br>Min. samples per leaf: 0.08<br>Min. sum of weights per leaf: 0.153<br>Max. features per split: 0.331<br>Max. sampling proportion: 0.774<br>Out-of-bag computation: False |
| | Performance | No feature engineering | Bagging | 0.678 ± 0.059 (0.711) | 0.472 ± 0.146 (0.452) | No. of weaker estimators: 352<br>Max. sampling proportion: 0.005<br>Max. features used: 0.104<br>Out-of-bag computation: True |
| | Security | No feature engineering | ElasticNet | 0.696 ± 0.228 (0.677) | 0.451 ± 0.167 (0.45) | Regularisation strength: 3.857<br>Mixing ratio between L1 and L2: 0.021 |

The shallow baseline classifier (first row of Table 15) achieved an accuracy of 0.5, whereas its precision, recall, and $F_1$-score values were all zero. The results are the same regardless of the FE method employed. Similar to our observations for RQ1, the deep Neural Network baseline classifier exhibited sensitivity to the preprocessing techniques applied, as seen in the second-to-last row of Table 15. We observe in Table 15 that Extreme Gradient Boosting is the strongest performer for RQ3, achieving consistent results in both accuracy (0.838) and $F_1$-score (0.825) across all FE techniques. A manual verification was undertaken to confirm the accuracy of these results to ensure that they did not stem from measurement errors. This inspection confirmed that there were minimal variations in performance between different FE techniques. For example, Extreme Gradient Boosting yielded an $F_1$-score of 0.824993 for non-engineered features and 0.824987 for normalised features, demonstrating a negligible difference of only 0.000006. The manual inspection also confirmed that log-transformation yielded slightly superior results (accuracy: 0.83817, $F_1$-score: 0.82505) compared to other FE techniques, so this combination is highlighted in **bold** in Table 15.

. These similarities highlight the model's robustness and consistency in predicting whether a user will cease participating. We therefore conclude that Extreme Gradient Boosting is a compelling choice for classifying user *Dropout*, particularly in scenarios where FE is not applicable or the feature distribution is unknown. We leveraged BO to identify the optimal hyperparameter settings that yielded the best results, followed by GA to ensure the robustness of these hyperparameters, ensuring that the hyperparameter values obtained from both BO and GA did not deviate by more than 5% (see Section 3.4.3). These results are depicted in Table 16.

As with RQ1 and RQ2, Kruskal-Wallis tests were used to check for differences in both accuracy and $F_1$-score across various model/FE combinations. The results revealed no statistically significant differences ($p > 0.45$ for both accuracy and $F_1$-score) across all model/FE combinations. Subsequent post-hoc tests were therefore not conducted.

While Extreme Gradient Boosting was the strongest and most consistent performer overall, several other models also exhibited consistency across multiple FE techniques. For example, Neural Network, K-Nearest Neighbours, and Random Forest consistently achieved $F_1$-scores exceeding 0.7 across all FE methods. The SGD classifier had the worst performance overall, which contrasts with its performance in RQ2, where SGD emerged as one of the top contenders. However, in RQ3, simpler models like Logistic Regression appear to be equally, if not more effective. This suggests that SGD may not be well-suited to the specific task of user *Dropout* classification in RQ3 and that its usefulness may be context-dependent.

As noted in Section 3.6, we re-evaluated the trained Extreme Gradient Boosting model's performance using a ground truth dataset of 385 unique randomly sampled users. The dataset was imbalanced, containing 124 positive class instances (i.e., users who did not drop out) and 261 negative class instances. The model achieved an accuracy of 0.881 and an $F_1$-score of 0.805 against our ground truth dataset.

### 4.4 Users' Textual-Based Dropout (RQ4)

This section presents results for RQ4: *How does a transformer-based model performs when compared to the chosen numerical model in predicting user dropout?*

Our fine-tuning of CodeBERT ran for 8 epochs, but experiments hinted that the model may have already converged at the $4^{th}$ or $5^{th}$ epoch, judging by the plateau of binary cross-entropy loss at 0.5. Fine-tuning took approximately 96 hours. Following this process, we noted that CodeBERT yielded a final accuracy of 0.822 and $F_1$-score of 0.809 for the validation set. To confirm the previous results from RQ3 (which involved analysing numeric-only data), we revisited the posts and code snippets written by the 385 users in the ground truth dataset used to benchmark RQ3 (see Section 4.3). The goal was to predict whether each user would dropout based on their posts and code snippets, rather than purely quantitative measurements. The posts and code snippets were preprocessed and tokenised as per the process outlined in Section 3.5.2. The fine-tuned CodeBERT model achieved an accuracy of 0.857 and an $F_1$-score of 0.792 against the ground truth dataset, which is comparable to the performance obtained with the Extreme Gradient Boosting model using numerical features. However, despite CodeBERT's ability to provide insights into dropout behaviour that numerical features might miss, it also has limitations in capturing certain dropout indicators that Extreme Gradient Boosting can identify. Table 17 presents the confusion matrices for both approaches.

Table 15. Benchmarking results for RQ3 (100 runs)

| Model Name | Log-transformation | | No feature engineering | | Normalisation | | Power-transformation | | Standardisation | |
|---|---|---|---|---|---|---|---|---|---|---|
| | **Accuracy** | **$F_1$-Score** | **Accuracy** | **$F_1$-Score** | **Accuracy** | **$F_1$-Score** | **Accuracy** | **$F_1$-Score** | **Accuracy** | **$F_1$-Score** |
| Shallow Baseline | 0.5 ± 0.0 (0.5) | 0.0 ± 0.0 (0.0) | 0.5 ± 0.0 (0.5) | 0.0 ± 0.0 (0.0) | 0.5 ± 0.0 (0.5) | 0.0 ± 0.0 (0.0) | 0.5 ± 0.0 (0.5) | 0.0 ± 0.0 (0.0) | 0.5 ± 0.0 (0.5) | 0.0 ± 0.0 (0.0) |
| AdaBoost | 0.656 ± 0.204 (0.632) | 0.678 ± 0.079 (0.684) | 0.66 ± 0.147 (0.632) | 0.669 ± 0.189 (0.657) | 0.678 ± 0.063 (0.707) | 0.679 ± 0.02 (0.674) | 0.693 ± 0.08 (0.707) | 0.65 ± 0.111 (0.67) | 0.693 ± 0.016 (0.711) | 0.648 ± 0.063 (0.674) |
| Bagging | 0.668 ± 0.273 (0.636) | 0.667 ± 0.174 (0.687) | 0.68 ± 0.039 (0.714) | 0.656 ± 0.224 (0.662) | 0.697 ± 0.088 (0.727) | 0.665 ± 0.114 (0.651) | 0.658 ± 0.086 (0.671) | 0.64 ± 0.165 (0.64) | 0.685 ± 0.006 (0.663) | 0.666 ± 0.201 (0.633) |
| C-SVM | 0.663 ± 0.172 (0.679) | 0.677 ± 0.226 (0.682) | 0.665 ± 0.21 (0.64) | 0.664 ± 0.241 (0.671) | 0.66 ± 0.044 (0.688) | 0.651 ± 0.02 (0.63) | 0.675 ± 0.177 (0.689) | 0.674 ± 0.078 (0.693) | 0.68 ± 0.031 (0.647) | 0.643 ± 0.058 (0.615) |
| Decision Tree | 0.752 ± 0.056 (0.764) | 0.742 ± 0.168 (0.759) | 0.74 ± 0.162 (0.746) | 0.729 ± 0.07 (0.761) | 0.724 ± 0.104 (0.725) | 0.712 ± 0.006 (0.731) | 0.747 ± 0.119 (0.78) | 0.735 ± 0.24 (0.765) | 0.757 ± 0.126 (0.788) | 0.746 ± 0.242 (0.709) |
| Extreme Gradient Boosting | **0.838 ± 0.224 (0.84)** | **0.825 ± 0.146 (0.856)** | 0.838 ± 0.031 (0.872) | 0.825 ± 0.095 (0.822) | 0.838 ± 0.084 (0.87) | 0.825 ± 0.182 (0.853) | 0.838 ± 0.237 (0.826) | 0.825 ± 0.171 (0.858) | 0.838 ± 0.134 (0.848) | 0.825 ± 0.297 (0.829) |
| K-Nearest Neighbours | 0.751 ± 0.206 (0.781) | 0.745 ± 0.186 (0.717) | 0.74 ± 0.225 (0.749) | 0.736 ± 0.006 (0.768) | 0.735 ± 0.25 (0.741) | 0.725 ± 0.249 (0.748) | 0.748 ± 0.005 (0.71) | 0.74 ± 0.047 (0.729) | 0.739 ± 0.297 (0.726) | 0.741 ± 0.021 (0.747) |
| Linear SVM | 0.713 ± 0.29 (0.722) | 0.703 ± 0.214 (0.706) | 0.708 ± 0.263 (0.691) | 0.7 ± 0.022 (0.693) | 0.663 ± 0.094 (0.679) | 0.656 ± 0.168 (0.684) | 0.719 ± 0.208 (0.708) | 0.708 ± 0.004 (0.738) | 0.688 ± 0.268 (0.716) | 0.703 ± 0.29 (0.736) |
| Logistic Regression | 0.71 ± 0.206 (0.684) | 0.699 ± 0.116 (0.68) | 0.634 ± 0.008 (0.628) | 0.657 ± 0.167 (0.687) | 0.643 ± 0.061 (0.62) | 0.627 ± 0.259 (0.634) | 0.714 ± 0.263 (0.736) | 0.702 ± 0.224 (0.733) | 0.645 ± 0.201 (0.624) | 0.627 ± 0.041 (0.631) |
| Random Forest | 0.744 ± 0.026 (0.739) | 0.736 ± 0.17 (0.712) | 0.743 ± 0.263 (0.752) | 0.734 ± 0.223 (0.734) | 0.742 ± 0.012 (0.776) | 0.734 ± 0.061 (0.706) | 0.744 ± 0.051 (0.756) | 0.734 ± 0.076 (0.707) | 0.743 ± 0.03 (0.715) | 0.735 ± 0.059 (0.7) |
| Ridge Classifier | 0.706 ± 0.125 (0.736) | 0.697 ± 0.018 (0.726) | 0.632 ± 0.059 (0.646) | 0.62 ± 0.032 (0.619) | 0.632 ± 0.168 (0.622) | 0.62 ± 0.036 (0.629) | 0.713 ± 0.042 (0.731) | 0.701 ± 0.013 (0.694) | 0.632 ± 0.24 (0.656) | 0.62 ± 0.068 (0.627) |
| SGD | 0.644 ± 0.126 (0.673) | 0.617 ± 0.174 (0.645) | 0.591 ± 0.208 (0.585) | 0.558 ± 0.072 (0.54) | 0.647 ± 0.287 (0.644) | 0.631 ± 0.291 (0.652) | 0.714 ± 0.16 (0.72) | 0.704 ± 0.254 (0.67) | 0.639 ± 0.095 (0.622) | 0.631 ± 0.148 (0.62) |
| Neural Network Baseline | 0.186 ± 0.005 (0.186) | 0.235 ± 0.013 (0.234) | 0.702 ± 0.04 (0.704) | 0.747 ± 0.045 (0.744) | 0.497 ± 0.031 (0.496) | 0.669 ± 0.036 (0.673) | 0.408 ± 0.024 (0.41) | 0.153 ± 0.009 (0.152) | 0.53 ± 0.03 (0.529) | 0.177 ± 0.009 (0.178) |
| Neural Network | 0.805 ± 0.124 (0.767) | 0.805 ± 0.192 (0.798) | 0.737 ± 0.015 (0.763) | 0.772 ± 0.079 (0.733) | 0.719 ± 0.208 (0.685) | 0.713 ± 0.157 (0.726) | 0.755 ± 0.124 (0.736) | 0.733 ± 0.277 (0.763) | 0.73 ± 0.161 (0.733) | 0.821 ± 0.02 (0.86) |

Values are presented as mean ± standard deviation (median)

Table 16. Optimal hyperparameter sets for log-transformation and Extreme Gradient Boosting

| **Hyperparameter** | **Value from BO** | **Value from GA** | **Difference (%)** |
|---|---|---|---|
| Base learner | Gradient boosting trees | Gradient boosting trees | - |
| Max. tree depth | 10 | 10 | 0 |
| Max. sampling proportion | 0.474 | 0.465 | 1.882 |
| Learning rate | 0.301 | 0.294 | 2.290 |
| No. of weaker estimators | 1,199 | 1,255 | 4.671 |
| L1 regularisation strength | 0.259 | 0.258 | 0.386 |
| L2 regularisation strength | 77 | 79 | 2.597 |

Table 17. Ground truth benchmark results

| **Numeric-based (Extreme Gradient Boosting)** | | |
|---|---|---|
| | Predicted Dropout | Predicted Non-Dropout |
| Actual Dropout | 244 | 18 |
| Actual Non-Dropout | 28 | 95 |
| **Textual-based (CodeBERT)** | | |
| | Predicted Dropout | Predicted Non-Dropout |
| Actual Dropout | 225 | 37 |
| Actual Non-Dropout | 18 | 105 |

From Table 17 we can observe that Extreme Gradient Boosting achieved a higher true negative rate (244) compared to CodeBERT (225). Conversely, CodeBERT exhibited a stronger performance in predicting true positives (105) as opposed to Extreme Gradient Boosting (95). For instance, Users.ID 9886516 was predicted to have dropped out based on numerical features alone, whereas the CodeBERT model analysing their posts and code snippets predicted otherwise. In this instance, CodeBERT's prediction aligned better with the ground truth data. On the other hand, Users.ID 4749866 was predicted to remain active (not drop out) by Extreme Gradient Boosting, while CodeBERT suggested dropout. In this scenario, the prediction from the numerical model aligned better with the ground truth data. These contrasting examples illustrate the potential benefits of a synergistic approach that leverages both code-based and numerical feature modelling, rather than relying solely on one method.

## 5 DISCUSSION AND IMPLICATIONS

### 5.1 Users' Answers (RQ1)

This section discusses the results obtained with respect to RQ1: *What is the best modelling approach to predict users' answers?*

Observing the results for RQ1, the highest performance was achieved by the Bagging ensemble model with standardisation technique, having an $R^2$ of 0.821 ± 0.098. It is important to distinguish Bagging from Random Forests as they are two completely different models, despite both being part of the ensemble family that create multiple decision trees. Unlike Random Forests, Bagging generates such trees using random subsets of the data with replacement (Guy et al., 2012). This process thus injects diversity into the model, leading to a reduction in variance and in the influence of any single data point (de Zarzà et al., 2023). Our results indicate that such processes can lead to improved model generalisability. Standardisation transforms features to have a zero mean and unit standard deviation (Shanker et al., 1996). This process improves the comparability of features, particularly when dealing with a dataset containing features measured in different units (Haga et al., 2019). In our study, we used diverse feature types, including metrics that depict less explicit contribution patterns (e.g., average *Post Polarity* and *Users' Contribution Frequency*), alongside directly quantifiable measures (e.g., total *Questions* and *Answers*, account age). *Post Polarity* measures the emotional orientation of user-generated content, while *User Contribution Frequency* quantifies the average number of *Questions* or *Answers* posted per day (Zolduoarrati et al., 2025c). Other examples include *User Popularity Index* which quantifies a user's fame within the community, while *Post Readability* evaluates the comprehensibility of their posts. Paired with simple quantifiable metrics derived from previous work (Zolduoarrati et al., 2024), these diverse facets collectively represent the construct of *contribution* which had remained abstract until recently (Zolduoarrati et al., 2025c). Our results demonstrate that standardising this heterogeneous feature set has yielded significant benefits. **Specifically, it is possible that Bagging reduces the variance of individual trees, and standardisation promotes fairer competition during splitting, leading to a more robust ensemble model**. This insight offers valuable implications for researchers and practitioners alike. For instance, those seeking to model contribution based on diverse feature sets similar to ours can use standardisation and Bagging techniques as a starting point, before iteratively refining their approaches. Starting with this foundation can reduce the time and effort required for unguided trial-and-error experimentation. Moreover, while decision trees (in the context of base estimators within Bagging ensembles) can already capture some non-linear relationships, standardisation can sometimes unlock additional non-linear patterns by transforming the features. Our results corroborate findings by Wang et al. (2011) where non-linear spectroscopy data is seen as a good fit for Bagging techniques.

In terms of the obtained optimal hyperparameters, the number of estimators identified through BO was 532. This indicates that for the Bagging model tasked with predicting users' *Answers*, **the ensemble should consist of approximately 532 individual decision trees, which is a relatively high number**. In fact, this number surpasses the maximum of 500 trees explored by Cuzzocrea et al. (2013) in their investigation of the impact of tree quantity on model performance. Moeini et al. (2024) also suggests that increasing the number of estimators in a Bagging model can generally lead to improved performance. The number of estimators is established as a critical hyperparameter in Bagging models, impacting their generalisation ability, as the model is constructed with a different number of weaker learners (Chowdhury et al., 2022). We also identified the optimal maximum sampling rate of 0.093, signifying that during training of each individual decision tree within the Bagging ensemble, 9.3% of the samples were drawn with replacement from the feature set. This coincides with the position of Liu et al. (2019), demonstrating that low sampling rates may outperform the conventional approach where sampling rate is set to 1 (i.e., all samples are used for bootstrapping). However, this number may be context-dependent (Sabzevari and Suárez, 2014), where different feature sets may yield different sampling rates. We also identified an optimal maximum feature inclusion ratio of 0.821, which means that during the training of each decision tree within the Bagging ensemble, approximately 12 features out of the total 15 (82.1%) were randomly selected for consideration. Previous research has established that increasing the inclusion ratio can generally improve the performance of Bagging ensembles (Shashank and Mahapatra, 2018), as a wider range of features at each node allows the trees to capture more complex relationships. **Our findings show that approximately 82% of all features would be more than sufficient in providing a good estimator**. More generally, around 82% of the variation in users' answer volume (i.e., number of questions a user is likely to answer) can be attributed to user-specific characteristics such as *Post Polarity*, *Post Readability*, and *User Popularity Index*. Our results led us to conclude that our trained Bagging estimator have adequately modelled users' total *Answers*, which is a variable commonly recognised as one of the most critical forms of contribution (Choetkiertikul et al., 2015; Mustafa et al., 2023; Zolduoarrati et al., 2024). These insights can translate to both further theoretical research and practical applications, discussed in Section 5.5.

In terms of other models, Neural Networks yielded lower results compared to Bagging despite the former being considered more complex (Omondiagbe et al., 2019). Granted, the Neural Network did not undergo the same level of hyperparameter tuning as the latter, a limitation that is discussed further in Section 6. The Epsilon SVM model exhibited the lowest performance, where its $R^2$ went as low as -0.255 ± 0.0 when paired with normalised features.

### 5.2 Users' Code Quality Violations (RQ2)

This section discusses the results obtained with respect to RQ2: *What is the best modelling approach to predict users' code quality?*

For RQ2, our analysis revealed that specific models yielded better performance compared to others. The most apparent was SGD which achieved top performance across five code quality prediction tasks: SQL Reliability, SQL Security, JavaScript Reliability, Python Reliability, and Ruby Performance. This strong performance aligns with prior research demonstrating SGD's ability to outperform similar shallow learning models like OLS regression and Random Forests in certain scenarios (Abida et al., 2024), especially large-scale, sparse problems in text classification and natural language processing (González-Briones et al., 2019). **Our findings add to this knowledge base, showing the generalisability of SGD which can also have good effect in predicting specific, more niche aspects of SE, namely code quality**. Moreover, SGD's observed success against these five tasks hints at the existence of hidden similarities in the way these violations appear in users' answers. In other words, the five *Code Quality*-related target variables on which SGD performs well may be more alike than we think. Alternatively, it could also indicate user tendencies to introduce certain types of violations that transcend languages. Future qualitative studies could be designed to explore these possibilities (see Section 7). The optimal hyperparameters for SGD also varied across all five target variables, suggesting that there is no single, one-size-fits-all set of hyperparameters that can effectively tackle all tasks. The difficulty of identifying general-purpose best hyperparameters is a well-documented challenge in the machine learning community (Yang et al., 2023). Regularisation strength, for example, ranged from 1.418 to 59.471, indicating that each *Code Quality* label may benefit from a different level of regularisation owing to the inherent data complexity (Reineking and Schröder, 2006). Target variables tuned with higher regularisation values (e.g., 59.471 for JavaScript reliability) may involve more complex data patterns that require stronger regularisation to prevent overfitting (Reineking and Schröder, 2006). In contrast, those with lower regularisation values (e.g., 1.418 for Python reliability) may have simpler data patterns for which overfitting is less of a concern.

The Bagging ensemble model achieved strong performance in predicting three *Code Quality* features: SQL Performance, JavaScript Readability, and Java Performance. These results suggest that performance and readability violations may be byproducts of complex, latent relationships between various users' features. One of Bagging's key strengths lies in leveraging diverse perspectives when combining forecasts from multiple decision trees (de Zarzà et al., 2023). **In other words, while other features likely influence these violations, their impact might not be directly linear or easily modelled by simpler algorithms, enabling Bagging to capture these intricate relationships via averaging predictions from multiple weaker estimators**. Unlike SGD, the optimal number of estimators in Bagging exhibited less variation across tasks, ranging from 352 trees for Java Performance to 569 for SQL Performance, with JavaScript readability falling at 417. These values are all relatively close, and interestingly similar to the number of trees found to be effective in RQ1 (532 trees) for modelling users' *Answers*. **We can therefore suggest that for Bagging models in this domain, a range of 400-500 trees may be more than sufficient to obtain good prediction results** regardless of the regression task the model is trained for. In addition, practitioners undergoing similar BO regimes may be able to leverage this insight as a starting point for tuning their own models. For instance, the hyperparameter search space for ensemble algorithms could be set to a narrower range, such as the 400-500 as we identified previously, instead of a wider interval.

Epsilon SVM was found to be most effective for modelling Python Readability, Python Security, and Ruby Security. This is an interesting juxtaposition, as Epsilon SVM had one of the worst scores in RQ1 for predicting users' *Answers*, yet did well when predicting *Code Quality* violations. Considering the model's inherent suitability for data with well-defined boundaries between positive and negative classes (e.g., secure vs. insecure snippets) (Ceperic et al., 2014), we may deduce that **security and readability violations might inherently exhibit clearer distinctions compared to other quality dimensions like performance or reliability, which can be more nuanced**. This finding aligns with terminology commonly used in SE literature, where studies pertaining to terms like *security hotspots* and *spaghetti code* suggest that security and readability violations are often easier to detect (Rafnsson et al., 2020). The underlying code security and readability characteristics may still exhibit some degree of complexity despite our outcomes favouring Epsilon SVM. To substantiate this point, experiments from BO revealed that the optimal kernel for Epsilon SVM was either the radial basis function (RBF) or second-degree polynomial, with both being commonly used in non-linear regression problems (Patle and Chouhan, 2013). Finally, BO also identified a maximum number of epochs of around 900 for all three security and readability tasks (878 for Python readability, 785 for Python security, and 918 for Ruby security). These values all fall below 1,000 epochs, hinting that for **similar BO applications in this domain, hard-limiting the number of epochs to 1,000 may already be a reasonable choice to avoid unnecessary training time**. Our results align well with the experiments by Chen et al. (2022) when evaluating the performance of K-Nearest Neighbours, Neural Network, AdaBoost, and SVM for cell classification. They showed that limiting the number of epochs to 1,000 iterations led to good results.

### 5.3 Users' Numeric-Based Dropout (RQ3)

This section discusses the results obtained with respect to RQ3: *What is the best modelling approach to predict users' dropout?*

For RQ3, the Extreme Gradient Boosting model with log-transformation FE achieved the highest performance, attaining a mean $F_1$-score of 0.825 ± 0.146. Our results corroborate prior research indicating the efficacy of Extreme Gradient Boosting for classification tasks compared to other models (Chen et al., 2023). In fact, the performance of Extreme Gradient Boosting is widely recognised within the machine learning community across domains, yielding consistent outcomes across a variety of domains, including public Kaggle competitions (Ibrahim et al., 2022)[27], geoscience applications like rock facies prediction (Zhang and Zhan), and econometric tasks such as crude oil price prediction (Gumus and Kiran, 2017). As a successor to Gradient Boosting (i.e., adding weak learners iteratively to address prior errors), **Extreme Gradient Boosting excels in efficiency via parallel tree generation and optimised data organisation (Malallah and Abdulrazzaq, 2023), which may explain its performance compared to their less innovative ensemble counterparts** (e.g., Random Forests or AdaBoost). Moreover, given user dropout is likely influenced by a complex interplay of factors (Mahbub et al., 2021), we posit that the model may be able to address this challenge by sequentially building decision trees where each tree focusses on improving upon the errors of the previous one. Our results also show that other models (e.g., Neural Networks and K-Nearest Neighbours) can also pose as alternatives in scenarios where Extreme Gradient Boosting might be less favourable due to external constraints, such as limitations imposed by the tech stack (given Extreme Gradient Boosting's reliance on the XGBoost third-party library[28]). Neural Network architectures require a careful balance of hidden layers to prevent both underfitting and overfitting (Uzair and Jamil, 2020). Three hidden layers have been empirically shown to be effective in many cases (Uzair and Jamil, 2020), and our results corroborate that this 'rule of thumb' is also suitable to predict user *Dropout*. For K-Nearest Neighbours, our experimental findings align with the work of Pal (2021), who conducted a benchmark study to select the best algorithm to classify skin diseases. They found that K-Nearest Neighbours can surpass traditional ensemble methods like Random Forest, similar to our findings.

[27] https://www.kaggle.com/competitions

**With respect to log-transformation, this approach normalises skewed features and compresses the value range, potentially improving their suitability for gradient boosting algorithms**. In fact, exploratory data analysis using goodness-of-fit tests revealed that several features fitted exponential distributions. For instance, the *YearlyDurationUsage* feature displayed a statistically significant departure from normality according to the Kolmogorov-Smirnov test (KS = 0.151, $p < 0.01$). Thus, corroborating Villadsen and Wulff (2021), we also show that the impact of log-transformation is not trivial. We posit that the ability of Extreme Gradient Boosting to handle latent complex relationships, paired with the improved normality and scaling achieved by log-transformation, enables more effective capturing of the patterns that influence user *Dropout*.

In terms of the hyperparameters, we first look at the number of weaker estimators (i.e., number of trees), maximum tree depth, and learning rate. These parameters are greatly influential on the model's complexity as they directly affect whether the model will overfit or underfit (Ibrahim et al., 2022). In our BO experiments, the optimal number of trees was selected to be 1,199, a relatively high number. The results of our tuning align with the experiments conducted by Vadhwani and Thako (2023), who found that the model's loss would largely decrease as the number of trees increases. **Thus, the relatively high number of trees likely contributes to the model's strong performance**. A larger maximum tree depth allows for more complex decision trees within the same model, but it can also lead to overfitting if not carefully chosen (Samat et al., 2020). Tuned using BO, we identified an optimal maximum tree depth of 10, which aligns with the work by Tarwidi et al. (2023), hinting that the optimal number of tree depths often falls within the range of 3 to 10. **Thus, such a number would likely balance performance and the risk of overfitting, as suggested by Lartey et al.** (2021). The learning rate controls the magnitude of updates during the tree boosting process, where smaller values have been shown to empirically improve performance (Ibrahim et al., 2022). BO identified an optimal learning rate of 0.301, aligning with the findings of prior works where **a similar learning rate yielded good accuracy and maintained satisfactory performance** (Wang et al., 2022).

Other hyperparameters can also play significant roles. For example, the optimal maximum sample proportion identified through BO was 0.474. This signifies that during the construction of each of the 1,199 trees in the ensemble, approximately 47.4% of the user data was sampled with replacement. This approach encourages each tree to learn from a slightly different subset of the data, leading to diverse estimations, which we surmise can contribute to the overall robustness of the model. Finally, our BO process identified L1 and L2 regularisation terms of 0.259 and 77, respectively. L1 regularisation promotes sparsity (Unser et al., 2016), which reduces the influence of certain features by applying a small penalty. L2 regularisation, on the other hand, penalises large weights, encouraging them to be smaller in magnitude (Ronao and Cho, 2016). Both help to control the overall complexity of the model, which is readily apparent in our results.

### 5.4 Users' Textual-Based Dropout (RQ4)

This section discusses the results obtained with respect to RQ4: *How does a transformer-based model performs when compared to the chosen numerical model in predicting user dropout?*

For RQ4, initial evaluation using the validation set indicated promising performance for CodeBERT with an accuracy of 0.822 and $F_1$-score of 0.809. To assess consistency and generalisability, the model was further evaluated on a ground truth dataset containing posts from 385 users, resulting in an accuracy of 0.857 and $F_1$-score of

[28] https://github.com/dmlc/xgboost

0.792. Our results further substantiate the findings of Zhou et al. (2021), demonstrating CodeBERT's generalisability to new data and tasks, particularly those beyond its original training data from the CodeSearchNet dataset (Feng et al., 2020). **CodeBERT-based solutions, given sufficient fine-tuning, can thus be seen to achieve competitive performance for tasks that were not in its original training regime**. As hinted in Section 4.4, our evaluation results against the 'ground truth' dataset reveal discrepancies between CodeBERT's text-based predictions and the numerical predictions from Extreme Gradient Boosting for user *Dropout*, yet this does not mean that one is better than the other. Indeed, Extreme Gradient Boosting were better suited to identify true dropouts (higher true negative rate of 244 compared to CodeBERT's 225), whilst CodeBERT excelled at correctly identifying non-dropout cases (105 true positives versus 95 for Extreme Gradient Boosting). Further manual investigation into these discrepancies revealed interesting insights. For example, Users.ID 10055307 had only one question and no answers. Based solely on numerical features like *Questions*, *Answers* and *User Contribution Frequency*, Extreme Gradient Boosting have classified this user as dropped out, which may be due to their similarity to other users who have dropped out. However, manual review of the user's single post (see Figure 6A) revealed it to be of high quality, that it adhered to Stack Overflow guidelines, and that it showcased prior research or testing, all indicative of a veteran user's habits (Pudipeddi et al., 2014). CodeBERT thus correctly classified this user as not dropout.

Conversely, some users were subject to accurate dropout predictions based on numerical features, while CodeBERT misclassified them as false positives. For example, Users.ID 3581678 posted only one question in 2014 (see Figure 6B) and has not logged in for three years. Their question received downvotes to -2 points despite being well-constructed and presenting reproducibility with sample data. CodeBERT therefore predicted continued engagement, presumably as the question's quality does not differ much from those presented by active users. **The unexplained downvotes on this well-constructed question could potentially contribute to a sense of discouragement or exclusion from the community, thus influencing the user's decision to drop out**. However, this remains an unproven hypothesis, and further qualitative investigation is necessary to determine the exact motivations behind both the downvotes and the user's dropout. The contrasting cases of Users.IDs 10055307 and 3581678 exemplify the complementary strengths and limitations of textual-based and numerical models. On one hand, Users.ID 10055307 highlights the value of textual analysis, as their informative post was overshadowed by the lack of activity on their part, leading to misclassification by numerical models. Conversely, Users.ID 3581678 demonstrates the effectiveness of numerical features (e.g., *User Popularity Index* and *Reputation*) at capturing user *Dropout* despite the transformer model's misprediction.

A similar phenomenon can be seen with Users.ID 623403. The Extreme Gradient Boosting model correctly predicted that they are likely to dropout (which they did) as their numerical variables shows relatively small participation (e.g., 5 answers and 11 questions). However, examination of one of their questions (Figure 7) reveals hallmarks of high-quality contribution, including precise problem formulation, contextual specificity, appropriate formatting, and demonstration of prior research and testing. Due to these qualitative characteristics, CodeBERT did not classify this user as dropout. The aforementioned examples highlights that the textual analysis through CodeBERT is able to capture semantic signals in users' posts and code snippets that might indicate continued engagement, aligning with one of the core strengths of transformer-based models; its ability to highlight relevant parts of the input regardless of their position in the sequence (Singh and Raman, 2024). Indeed, **textual-based models must be supplemented by numeric-based ones to provide a robust ensemble, as textual data may be able to reveal patterns missed by numerical measurements (and vice versa)**. Combining textual and numerical features has been shown to improve prediction robustness (Cerchiello et al., 2018). For instance, integrating financial news was shown to introduce latent information that was otherwise intangible in standard numerical variables (Cerchiello et al., 2018). In the context of our study, our results bridge gaps in prior predictive modelling studies, which have typically relied on either numeric features or textual analysis in isolation (Olatinwo and Demmans, 2024; Shah and Pomerantz, 2010).

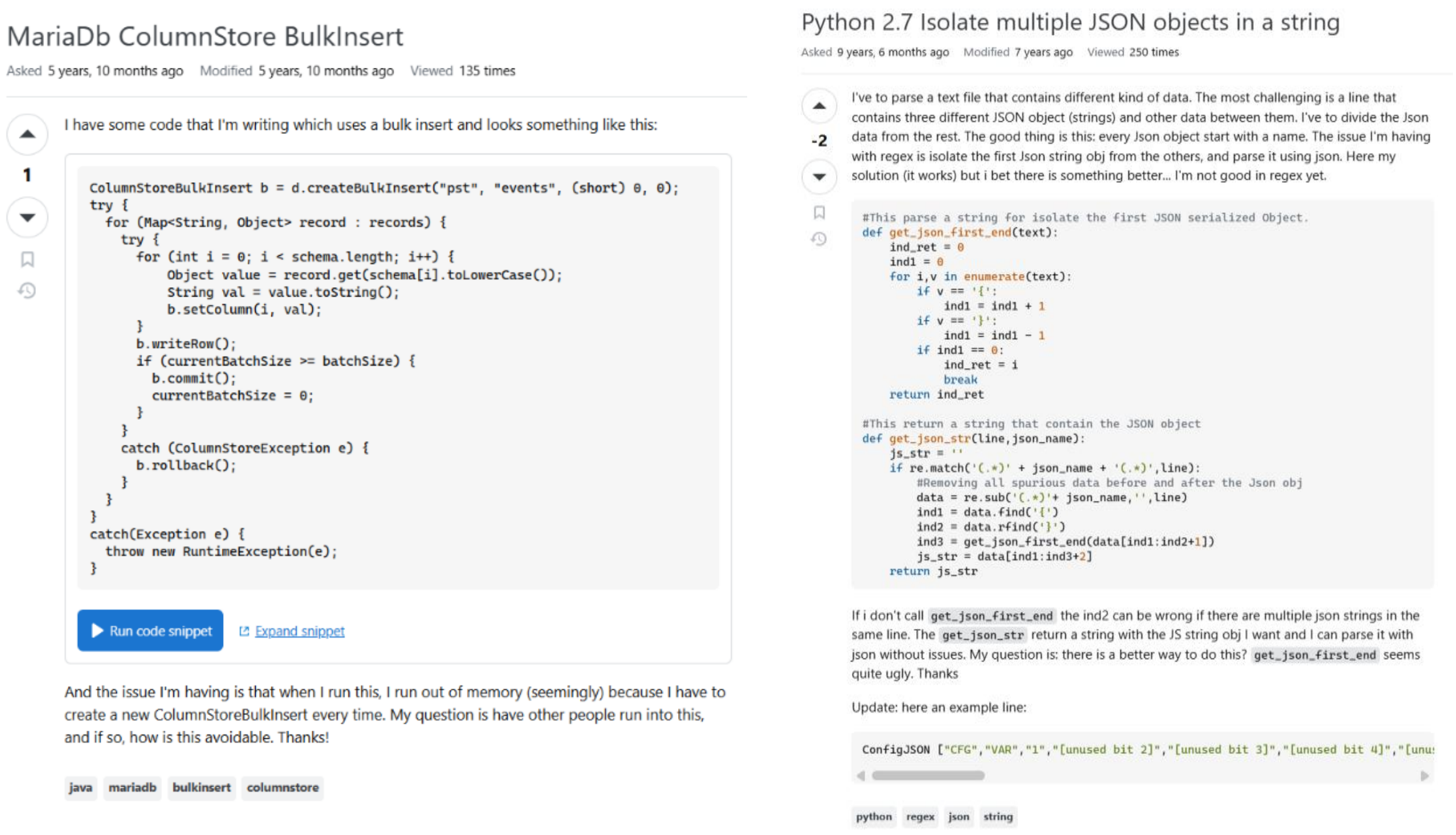

(A) (B)

Figure 6. Example posts by Users.ID 10055307 (A) and Users.ID 3581678 (B)

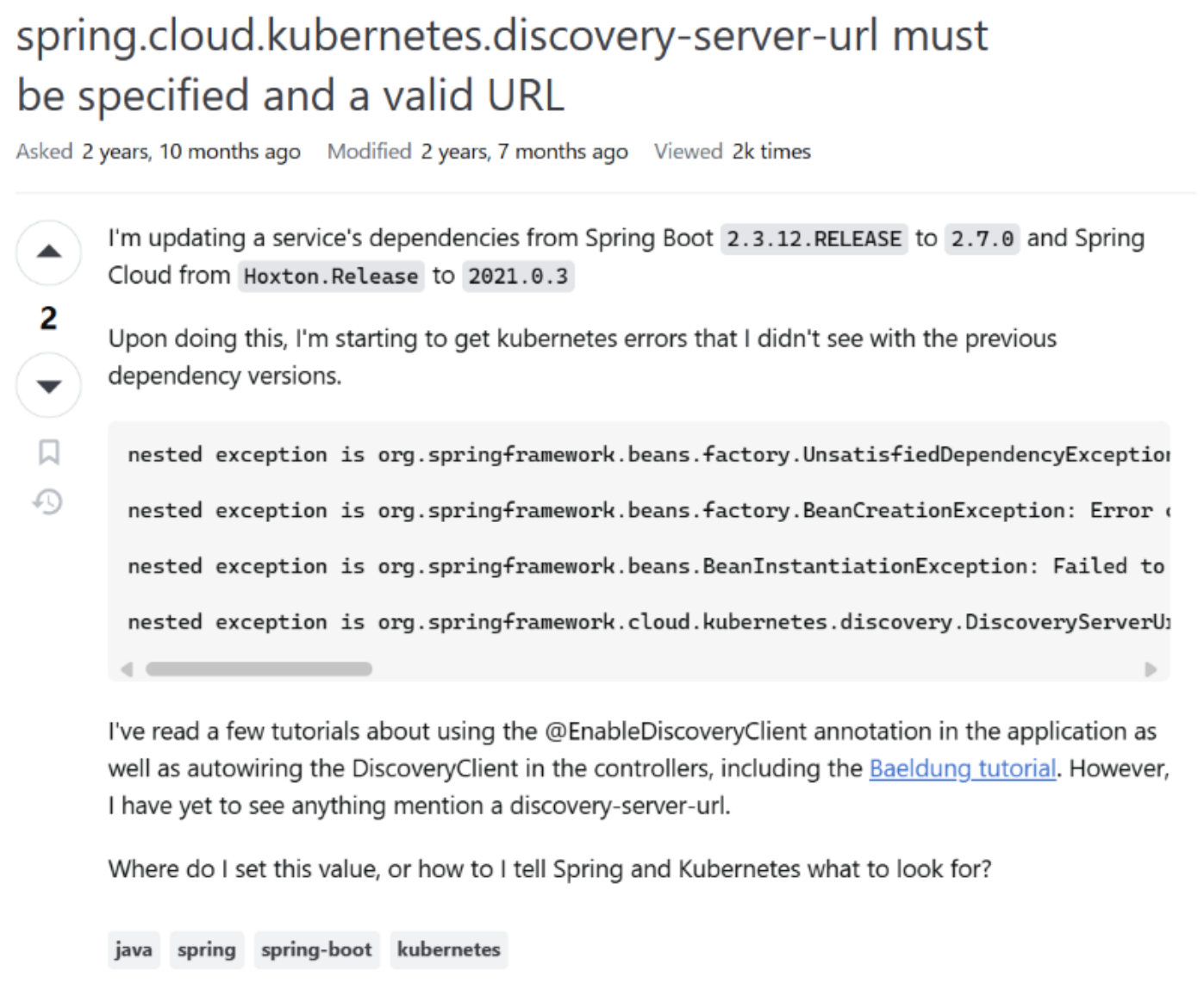

Figure 7. Example question by Users.ID 623403

CodeBERT therefore validates the numerical model by demonstrating that dropout prediction is not merely an artifact of the chosen methodology; but rather represents a genuine behavioural pattern detectable through multiple analytical lenses. Deeper qualitative exploration can be done to ascertain the true nature of patterns indicative of user *Dropout* (as discussed in Section 7).

## 5.5 Implications

Our work builds on prior work aimed at predicting user-inherent variables on CQA portals (as discussed in Section 2) by introducing a novel approach that incorporates a wider range of metrics and models. In doing so, we have provided several implications for theory, practice, and policy development. From a theoretical perspective, our work addresses an epistemological gap within the literature, where previous predictive endeavours have typically employed only 3-5 models or utilised arbitrary model selection when predicting Stack Overflow user contributions. We recommend that researchers adopt our model pool, or even extend them, when seeking to predict similar dimensions in CQA settings. Moreover, our findings suggest the need for an integrated predictive framework that combines both numeric and text-based approaches for modelling complex user behaviours – not only user dropout – in online SE communities. The complementary nature of Extreme Gradient Boosting for numeric and CodeBERT for textual features demonstrates that unimodal approaches are insufficient for capturing the otherwise multifaceted nature of user behaviour. In fact, the usage of CodeBERT represents a leap within the current research niche, as previous research using CodeBERT on Stack Overflow data has not been used to predict user dropout (Zhang et al., 2022).

For SE practice, our results provide guidance for implementing early intervention systems within CQA platforms. Platform administrators are advised to deploy ensemble prediction models that incorporate both quantitative metrics and semantic text analysis to identify 'at-risk' users before they disengage, or perhaps identify latent factors that inhibit users from actively participating. The implementation of certain models can also be developed to maintain CQA sites' relevance. For example, implementing the Bagging ensemble model to predict answers (as seen in RQ1) may allow platform designers to identify 'rising stars' i.e., potential high-value contributors. Another example is employing Extreme Gradient Boosting to trigger automated retention mechanisms when dropout probability exceeds predetermined thresholds. Furthermore, development teams are recommended to integrate our fine-tuned CodeBERT model into code review processes to proactively identify code quality issues that numeric metrics alone might miss, particularly for security and readability violations, as both exhibit clearer distinctions compared to performance or reliability dimensions. Additionally, by analysing historical data on downvoted questions and answers, our fine-tuned CodeBERT model may be harnessed to generate examples of well-structured and informative questions and answers. This educational resource could be presented to new users to help them avoid downvotes and navigate the platform more effectively.

Regarding policy development, governance policies should incorporate educational interventions when users exhibit patterns associated with potential dropout or poor code quality contributions – which can be characterised using our fine-tuned CodeBERT model. These policies should also include mechanisms to systematically collect cases where numeric and text-based predictions diverge (as exemplified by Figures 6 and 7), as these divergences represent critical insights as to why contributors with good quality submissions tend to dropout, or why established users unknowingly downvote newcomers despite their compliance with community guidelines.

## 6 THREATS TO VALIDITY

This section outlines the threats to our work, which are considered in relation to internal, external, and construct validity.

### 6.1 Internal Validity

The first internal threat to validity concerns the scope of the training data. Our models were trained and evaluated on a subset of Stack Overflow data for users whose locations were identified as being within the US, as previously operationalised (Zolduoarrati et al., 2025b, 2024, 2025c). This raises the possibility that using a larger and more diverse dataset for training could potentially improve model performance. However, rigorous Bayesian optimisation was conducted to identify the most optimal hyperparameter configurations, followed by a validation step using genetic algorithms. This two-pronged approach provides confidence that the model configurations we obtained represent the best possible performance achievable with the given data. Thus, we believe the performance threat was reduced. In terms of our chosen models, despite having a pool of 21 algorithms, we evaluated only 18 models for regression tasks (RQ1 and RQ2) and 12 models for classification (RQ3). Our selective application reflects the inherent functional constraints of certain algorithms, as imposing inappropriate tasks for models that were not designed for such tasks (e.g., using classification-only models for regression tasks) would compromise methodological validity. Indeed, this approach may have precluded universal conclusions across all 21 algorithms due to epistemological limitations. However, it enhances scientific integrity by ensuring each model operates within its intended functional domain. Consequently, all reported results in this manuscript stem from the appropriate application of algorithms to their specifically designed tasks, minimising erroneous conclusions. Our chosen imputation strategies for each variable's missing values may also not be optimal. For instance, zero imputation was used for the *Answers* variable, yet alternative strategies (e.g., EM or K-Nearest Neighbours) can still improve model accuracy. To mitigate this limitation, selection of the most appropriate imputation strategy for each variable (see Table 3) was grounded in sound reasoning and a deliberate rational approach following numerous internal discussions between the researchers.

Our use of Bonferroni correction for Conover-Iman tests represents another threat. This conservative approach minimises false positives (type I errors) yet can also mask true effects (type II errors) by reducing statistical power. While Bonferroni correction ensures robust findings, alternative methods like Holm or Benjamini-Hochberg corrections might reveal more effects at the cost of slightly increased risk for type I errors (Vasilopoulos et al., 2016). We concede this risk, as we wanted to avoid overly speculative claims.

Next, our study utilised a specific set of features to address RQ1, RQ2, and RQ3. However, features that were not accounted for in our study can serve as strong predictors for users' *Answers*, *Code Quality* violations, and their *Dropout* classification. In fact, the dependent variables of subsequent RQs could potentially serve as strong predictors for earlier ones. For instance, *Code Quality* violations (RQ2) and user *Dropout* (RQ3) might be strong predictors for users' *Answers*. This limitation is inherent to the chosen sequential study design, which was adopted to progressively increase the complexity of each RQ and the associated predictor variables.

Our study used four FE approaches in conjunction with various models. While this represents a significant exploration of feature space, we acknowledge that other relevant FE techniques might have been overlooked. Moreover, there might be other valuable models beyond those explored in this study. However, we employed a rigorous evaluation process to ensure the chosen models were well-suited to our research objectives, as outlined in Section 3.4, thus minimising this threat. We also would like to emphasise that to the best of our knowledge our study is the first to comprehensively evaluate the impact of four distinct FE methods on a wide array of models for predicting user-related variables (*Answers*, *Code Quality*, and *Dropout*) on Stack Overflow, before triangulating our findings with transformer-based models. Therefore, in spite of the potential for unexplored FE and modelling approaches, we believe our work is extensive and provides valuable insights.

In terms of our benchmarks, while the experiments showed that shallow learning algorithms largely outperformed Neural Networks, this might not be a true reflection of their capabilities. The primary reason may lie in the hyperparameter tuning process where the Neural Networks were not subject to a similar level of tuning effort as their shallow learning counterparts. This is because tuning Neural Networks requires extensive resources to achieve optimal performance, which can be computationally expensive. As a result, the observed performance gap may be an artifact of this bias, rather than a true reflection of their inherent capabilities. Some models were also intentionally omitted from our benchmarks due to their obsolescence (e.g., gradient boosting) or the availability of objectively superior alternatives (e.g., ElasticNet over Lasso and Ridge). We acknowledge the possibility that some of these excluded models could have yielded superior results, but our literature review strongly suggested that their performance would likely be inferior to their more advanced counterparts. These models were deliberately excluded to optimise resource allocation for our study. We therefore have minimised this threat to the fullest possible extent.

All 20 regression tasks associated with RQ2 yielded $R^2$ values above 0.65. Generally, computational sciences consider an $R^2$ value of 0.70 or higher as the ideal benchmark (Santos et al., 2022), yet the inherent complexity and variability of human behaviour poses significant challenges for empirically modelling (Schroeder, 2009). Consequently, the obtained $R^2$ values, while not exceptional, are nevertheless considered satisfactory given the abstract nature of the phenomena under investigation.

With respect to CodeBERT, one significant constraint is its effective maximum limit of 509 tokens (see Section 3.5.2). This means that any textual sequence exceeding this length will be truncated. This threat is evident in works that have used CodeBERT, but it is likely to exert minimal impact (Mashhadi and Hemmati, 2021). We have also used CodeBERT to validate the findings of RQ3, which pertains to user *Dropout* classification. Given the prevalent use of transformer-based models in classification tasks, it was deemed appropriate to validate only the results of RQ3. In contrast, RQ1 and RQ2 addressed regression tasks, for which the application of transformer models is less established in literature. Thus, while RQ1 and RQ2 remain unvalidated, the successful validation of RQ3 using CodeBERT suggests that the findings of RQ1 and RQ2 are likely to also be accurate, as RQ3 builds progressively on RQ1 and RQ2. Another limitation is that models trained exclusively on natural language, such as T5 or SpanBERT, may potentially outperform CodeBERT. However, CodeBERT has emerged as the *de facto* standard for evaluating both natural and programming languages, consistently demonstrating strong performance in numerous studies (see Section 3.5.1). Our selection of CodeBERT is rooted in our decision to prioritise the use of a well-established and reliable tool, mitigating the risk of suboptimal results from models that are less established.

We also opted not to employ state-of-the-art, newer LLMs due to the significant computational costs and time investment associated with their fine-tuning (Bowen et al., 2024). While it is highly likely that contemporary generative LLMs can perform better due to improved reasoning capabilities, at the same time there is no guarantee that these models would also outperform our fine-tuned CodeBERT. In fact, research by Liu et al. (2023) which was conducted within a similar timeframe as our study demonstrated that code synthesis frequently contained subtle errors. Moreover, at the time of writing, LLMs were relatively new, unpredictable, and resource-intensive (Roberts et al., 2024). Their integration would have necessitated extensive fine-tuning protocols for code assessment purposes, constituting an independent research initiative in its own right. We acknowledge this limitation as a potential threat to our study's validity. Finally, some models such as Random Forests incorporate elements of randomness. To minimise the potential threat to reproducibility, we consistently set the random seed to 42 on applicable models, following established practices (Arunachalaeshwaran et al., 2022).

### 6.2 External Validity

The external validity of our findings may be limited by the scope of the training data, as our models were exclusively trained on a dataset of 760,809 user records from the US, representing only 4.25% of Stack Overflow's global user base (17,922,426) at the time of data collection. This raises a threat related to the generalisability of the models to the broader international Stack Overflow community, as well as their applicability to other question-and-answer platforms like Cross Validated[29] or Quora[30]. Users from different geographical locations might exhibit distinct behavioural patterns that our models were not equipped to capture. Nonetheless, we found that our findings largely align with other literature, demonstrating expected behaviour in terms of results, performance, and hyperparameter values. This alignment validates our pre-trained models despite the relatively small training data.

As outlined in Section 6.1, our evaluation pool encompassed 18 models for regression (RQ1 and RQ2) and 12 models for classification (RQ3). This differentiated implementation stems from the inherent task-specificity of certain algorithms. We concur that this approach introduces a threat on cross-algorithmic generalisability across our total pool of 21 algorithms, yet it was methodologically necessary to maintain scientific integrity by applying each algorithm exclusively within its appropriate domain. Furthermore, this limitation is mitigated given that overarching model families were sufficiently represented across both regression and classification tasks. For example, while specific implementations such as C-SVM and Epsilon-SVM are restricted to classification and regression applications respectively, both were derived from the same overarching Support Vector Machine model family. Thus, our findings maintain a certain degree of cross-domain applicability, minimising generalisability threats to the fullest extent feasible whilst maintaining methodological rigour. To answer RQ4, a pre-trained CodeBERT model was fine-tuned based on the February 12, 2022 repository commit, which was the latest at the time of writing. However, later versions of CodeBERT may have undergone improvements that could produce better performance than the version we employed in our study. Thus, the findings of this study are specific to the particular version of CodeBERT used. CodeBERT also represents a transformer-based architecture rather than contemporary LLMs. The generalisability of our results to recent LLMs such as Gemma, Gemini, and GPT-4o may be limited, as these models were in early developmental stages during our study period, exhibiting variable performance while imposing significant financial and computational costs (Roberts et al., 2024). Nevertheless, exploring the utility of these models for the tasks investigated in this study remains a promising direction for future work (see Section 7).

Finally, Stack Overflow is a dynamic platform where users' contribution patterns and their code quality can evolve over time due to changes in platform features and emergent technologies (e.g., generative models like ChatGPT). Our study was conducted during April to December 2024. During this specific point in time, we may not capture the full spectrum of user behaviour as the SE community continues to advance. For instance, the growing popularity of generative LLMs has been shown to alter user participation patterns (Kabir et al., 2024), yet these LLMs might not significantly impact the core value of Stack Overflow as these generative models often generate inaccurate output (Kabir et al., 2024). Stack Overflow, on the other hand, provides highly tailored and accurate solutions to specific problems. In fact, Stack Overflow recently implemented a new rule prohibiting answers written by generative models[31], which accentuates that the platform maintains its relevance as a trusted source of reliable technical information.

[29] https://stats.stackexchange.com

[30] https://www.quora.com

[31] https://meta.stackoverflow.com/questions/421831

### 6.3 Construct Validity

The chosen features may not be the most relevant ones for predicting the target variables. For instance, it might not be immediately obvious why a feature like "*Average AboutMe Polarity*" can be directly related to a user's total *Answers* (RQ1). In other words, the relationship between two variables might not appear logical at first glance. However, we leveraged features validated in prior works (Zolduoarrati et al., 2025b, 2024, 2025c) to predict user behaviour on Stack Overflow, thereby ensuring that all relevant features are included as potential predictors. Additionally, we addressed potential multicollinearity by calculating VIFs for each feature, further mitigating this concern.

There are also limitations in terms of the target variables. First, the definition for code quality violations (i.e., target variable of RQ2) only encompasses violations with respect to five programming languages (SQL, JavaScript, Python, Ruby, Java) and four quality dimensions (reliability, readability, performance, security). This restricted scope might not comprehensively represent the entire construct of *code quality*, as there are certainly other programming languages that were not taken into account (e.g., Go, TypeScript, and Rust). We concede that ideally, a more extensive range of languages and quality dimensions could be considered, yet our current scope draws upon prior work where these code quality facets had undergone investigation in terms of representativeness (Zolduoarrati et al., 2025b). Moreover, our four dimensions align with those identified by Ndukwe et al. (2023): *Functionality*, *Readability*, *Efficiency*, *Security*, and *Reliability*. We consider *Functionality* to be a subset of *Reliability* as reliable code is inherently functional. *Efficiency* is also synonymous with *Performance* as performant code is expected to execute tasks with minimal computational cost. Thus, despite the presence of this threat, we are certain our operationalisation of *code quality* is valid. Similarly, the definition of user dropout (the target variable for RQ3) warrants further consideration, given that the concept of *user dropout* lacks a universally agreed-upon definition. Some scholars, like Slag et al. (2015) define dropout in terms of *one-day-flies*, i.e., users who contribute only a single post before becoming inactive. This definition differs from the one adopted in our study, where dropouts are defined as users who have not participated in discussions for 60 days, derived from prior work (Zolduoarrati et al., 2025c). This inconsistency across semantics poses a threat, as the observed results might be specific to the chosen definition and may not be applicable to a broader construct for the same term.

## 7 CONCLUSION AND FUTURE WORK

Previous studies that tackled prediction using Stack Overflow data tended to employ limited benchmarks of 3-5 models or adopted arbitrary selection methods. Our study offers an advancement to the current state of knowledge, where we benchmarked a comprehensive suite of 21 algorithms, encompassing classic linear and Bayesian models, SVM sub-variants, ensemble methods, and feed-forward neural networks. This comprehensive set is composed of 9 regression-only models, 3 classification-only models, and 9 hybrid models capable of performing both regression and classification tasks. We evaluated the effectiveness of these approaches at predicting users' *Answers*, *Code Quality* violations, and user *Dropout*. Four feature engineering techniques were conducted prior to each model's execution to determine which model and feature engineering combination would prove the most effective for the given task. Each model's hyperparameters were first subjected to Bayesian optimisation where genetic algorithms were employed as a validation step to confirm the correctness of the configurations. Finally, we leveraged CodeBERT – a prominent transformer model adept at handling both natural and programming languages – to classify user dropout based on their posts (both questions and answers) and code snippets. This final step triangulated the findings from the preceding quantitative models, and also enabled us to explore latent insights into user dropout that are otherwise not captured by numerical measurements.

Our benchmark highlighted optimal combinations of model and feature engineering for each task. Predicting how many questions a user is likely to answer proved most successful with a bagged ensemble model paired with standardisation techniques. However, predicting code quality across various languages and quality dimensions necessitated a more diverse set of configurations, though models like SGD, Epsilon SVM, and Bagging ensembles emerged as top performers in this domain. Finally, to predict user dropout, Extreme Gradient Boosting coupled with log-transformation emerged as the most effective numeric-based approach. CodeBERT triangulated the quantitative findings by showing that some predictions were overlooked by numeric models, such as users who have not dropped out but have posted only one well-structured question. CodeBERT was able to effectively classify these users, which may stem from those users' textual patterns aligning with those of other active users. Numeric models likely missed this due to the lack of answer data. These observations highlight the importance of combining textual-based and numerical-based models into an ensemble approach. We believe the usage of both approaches fosters robustness given that textual models were able to capture patterns that eluded numerical analysis, and vice versa.

This research lays the groundwork for promising future explorations. One potential avenue involves evaluating the efficacy of alternative models not explored within the scope of this study. Additionally, further research may delve into extensive hyperparameter tuning for the Neural Network, aiming to maximise its performance and properly determine whether it is comparable to other shallow learning models. Furthermore, scholars are encouraged to address limitations in our operationalisations, expanding the constructs of *code quality* and *user dropout* to provide a more comprehensive understanding of Stack Overflow's and other CQAs underlying dynamics. A benchmarking study comparing various LLMs in predicting answers, code quality, and user dropout on Stack Overflow is also a worthwhile endeavour to triangulate our findings. Such an analysis has become more accessible as computational efficiency and fine-tuning affordability improve, evidenced by recent advances in LLMs (e.g., DeepSeek-R1). Finally, we strongly advocate for the validation of our findings with qualitative methods. Examples include interviews to explore the reasons behind downvotes on well-constructed questions, and surveys to investigate factors influencing the pervasiveness of specific code quality violations. These qualitative approaches can illuminate latent user experiences that transcend the quantitative patterns observed in this study.

## 8 DATA AVAILABILITY

Our replication package is available online at https://zenodo.org/records/15330992, ensuring the study's reproducibility and transparency. This encompasses a suite of methodological demonstrations, including the evaluation of Guesslang, the application of regular expressions, and other detailed expositions behind the selection of benchmarked algorithms, the treatment of missing data, and relevant underlying design decisions. Furthermore, the replication package includes the complete set of empirical results, which exceeded the permissible length for inclusion within this manuscript.

## 9 DECLARATIONS

The researchers declare no competing financial interests or personal relationships that could have influenced the work presented in this manuscript.

# APPENDIX A

Table A1. VIF values for RQ1 features

| Feature | VIF |
|---|---|
| *Post Attention to Detail* | 29.172 |
| *Post Readability* | 21.041 |
| *Badges* | 10.332 |
| *User Contribution Frequency* | 8.216 |
| *Reputation* | 7.188 |
| *YearlyDurationUsage* | 4.350 |
| *User Profile Completion Rate* | 3.213 |
| *Questions* | 3.063 |
| *Comments* | 2.512 |
| *Edits* | 2.093 |
| *Average AboutMe Polarity* | 1.865 |
| *ProfileLength* | 1.840 |
| *Views* | 1.787 |
| *UpVotes* | 1.772 |
| *User Popularity Index* | 1.730 |
| *Comment Polarity* | 1.558 |
| *Answer Polarity* | 1.452 |
| *Question Polarity* | 1.357 |
| *Code Length* | 1.347 |
| *DownVotes* | 1.257 |

Table A2. VIF values for RQ2 features

| Feature | VIF |
|---|---|
| *Post Attention to Detail* | 29.224 |
| *Post Readability* | 21.042 |
| *Badges* | 11.069 |
| *Reputation* | 8.929 |
| *User Contribution Frequency* | 8.365 |
| *YearlyDurationUsage* | 4.357 |
| *Answers* | 4.024 |
| *User Profile Completion Rate* | 3.214 |
| *Questions* | 3.147 |
| *Comments* | 3.058 |
| *Edits* | 2.173 |
| *Views* | 1.878 |
| *Average AboutMe Polarity* | 1.865 |
| *ProfileLength* | 1.840 |
| *UpVotes* | 1.778 |
| *User Popularity Index* | 1.775 |
| *Comment Polarity* | 1.558 |
| *Answer Polarity* | 1.452 |
| *Question Polarity* | 1.357 |
| *Code Length* | 1.347 |
| *DownVotes* | 1.290 |

Table A3. VIF values for RQ3 features

| Feature | VIF | Feature | VIF |
|---|---|---|---|
| *Post Attention to Detail* | 29.310 | *UpVotes* | 1.762 |
| *Post Readability* | 21.049 | *Comment Polarity* | 1.558 |
| *Badges* | 11.354 | *Answer Polarity* | 1.461 |
| *Reputation* | 8.948 | *Question Polarity* | 1.357 |
| *Java Avg. Performance Violation Density* | 8.863 | *Code Length* | 1.348 |
| *User Contribution Frequency* | 8.584 | *SQL Avg. Readability Violation Density* | 1.329 |
| *Java Avg. Security Violation Density* | 8.574 | *DownVotes* | 1.315 |
| *YearlyDurationUsage* | 4.376 | *Python Avg. Readability Violation Density* | 1.256 |
| *Answers* | 4.073 | *Python Avg. Reliability Violation Density* | 1.245 |
| *SQL Avg. Security Violation Density* | 3.386 | *Ruby Avg. Readability Violation Density* | 1.244 |
| *SQL Avg. Performance Violation Density* | 3.322 | *Ruby Avg. Reliability Violation Density* | 1.207 |
| *User Profile Completion Rate* | 3.219 | *JavaScript Avg. Reliability Violation Density* | 1.195 |
| *Questions* | 3.158 | *JavaScript Avg. Readability Violation Density* | 1.093 |
| *Comments* | 2.972 | *JavaScript Avg. Security Violation Density* | 1.057 |
| *Edits* | 2.145 | *Python Avg. Security Violation Density* | 1.032 |
| *Views* | 1.866 | *Python Avg. Performance Violation Density* | 1.018 |
| *Avg. AboutMe Polarity* | 1.865 | *Ruby Avg. Security Violation Density* | 1.012 |
| *ProfileLength* | 1.840 | *JavaScript Avg. Performance Violation Density* | 1.010 |
| *Java Avg. Reliability Violation Density* | 1.839 | *SQL Avg. Reliability Violation Density* | 1.006 |
| *Java Avg. Readability Violation Density* | 1.802 | *Ruby Avg. Performance Violation Density* | 1.002 |
| *User Popularity Index* | 1.776 | | |